%% file: body.tex
\documentclass[twocolumn]{revtex4-1}
\usepackage{graphics}
\usepackage{graphicx}
\usepackage{amsfonts}
\usepackage{epsfig}
\usepackage{color}
\usepackage{amsmath}
\usepackage{amssymb}
\usepackage{mathtools}
\usepackage{subfigure}
\usepackage{caption}
\usepackage{geometry}
\usepackage{layout}
\usepackage{tabularx}
\usepackage{tikz}
\usepackage[normalem]{ulem}
\usetikzlibrary{shapes.callouts}
\usetikzlibrary{calc,arrows,decorations.pathmorphing,intersections,patterns}
\tikzset{
  level/.style   = {very thick, black },
  init sublevel/.style={black,dashed},
  ionization/.style={black,dashed},
  transition/.style={black,->,>=stealth',shorten >=1pt},
  connect/.style = { dashed, red },
  notice/.style  = { draw, rectangle callout, callout relative pointer={#1} },
  label/.style   = { text width=2cm }
}
\usepackage{multirow}
\begin{document}

\title[]{A global model of cylindrical and coaxial surface-wave discharges}
\author{Efe Kemaneci$^1$, Felix Mitschker$^2$, Marcel Rudolph$^2$,
Daniel Szeremley$^1$, Denis Eremin$^1$, Peter Awakowicz$^2$, Ralf Peter Brinkmann$^1$}

\affiliation{$^1$ Institute for Theoretical Electrical Engineering, Ruhr-Universit{\"a}t Bochum, 
Germany}
\affiliation{$^2$ Institute for Electrical Engineering and Plasma Technology, Ruhr-Universit{\"a}t Bochum, 
Germany}

\bibliographystyle{unsrt}

\begin{abstract}
\noindent 
A volume-averaged global model is developed to investigate surface-wave 
discharges inside either cylindrical or coaxial structures. The neutral and 
ion wall flux is self-consistently estimated based on a simplified analytical 
description both for electropositive and electronegative plasmas. The simulation 
results are compared with experimental data from various discharge setups of 
either argon or oxygen, measured or obtained from literature, for a continuous 
and a pulse-modulated power input. A good agreement is observed between the 
simulations and the measurements. The calculations show that the wall flux often 
substantially contributes to the net loss rates of the individual species. 

\end{abstract}

\maketitle

\input{intro}
\input{setup}

\input{phys_mod}

\input{globalmodel}
\input{results}
\input{conc}
\input{appendix}

\input{chemkin}

\newpage
\bibliographystyle{unsrt}
\bibliography{refdrakachem}

\end{document}

%% file: intro.tex
\section{Introduction}

A unique structure of microwave induced surface-wave discharges allows axial 
wave propagation along the dielectric-plasma interface and this axially separates them 
from other types of processing plasmas. The applicability 
for a large substrate area and the absence of electrode degradation or pollution 
makes them attractive for a variety of purposes. A computational investigation 
of these plasmas with a feeding gas of either argon or oxygen is the main 
objective of this study with a focus on a coaxial plasmaline 
\cite{Ste13,Bah13_2,Nav16} used in the processing of the PET bottles. Considered 
operation conditions cover a pressure range of $25-8000\:$Pa in cylindrical 
(surfatron) or coaxial (plasmaline) structures with either continuous or 
pulse-modulated power input. 

One-dimensional models are used for the axial \cite{Gla80,Pet99,Mak01} as well 
as radial characteristics \cite{Fer83} and complemented with a series of 
zero-dimensional kinetic models \cite{Fer91,Kut10} to provide a wider insight 
into various aspects. A variety of more detailed models are developed for a 
feeding gas of argon \cite{Alv09,Rah14_1,Jim12}, hydrogen \cite{Obr15} and 
oxygen \cite{Kem15} that are often associated with numerically expensive 
simulations. A computationally efficient zero-dimensional alternative is a 
transient global model that additionally averages over the position variable of 
the phase space. 
Although, an agreement is obtained with the measurements in surfatron oxygen
plasma as well as with spatially resolved models in a microwave reactor \cite{Kem15},
they are rarely implemented in the considered plasmas. This implementation also requires
a self-consistent estimation of the ion wall flux that is still missing for the  
surface-wave discharges.

The global models are mostly implemented in comparatively very low pressures 
\cite{ChabertBook}, probing a variety of aspects \cite{Kem14,Kem16}, where the 
ion wall flux rates are analytically calculated based on Bohm point 
edge-to-centre ratios. An edge-to-centre ratio at the lower edge of the 
low-pressure regime is analytically calculated for argon \cite{God86} and later 
derived for larger pressures with a proper combination \cite{LieBook2005}.
The derivations are heuristically generalized for the electronegative plasmas
in the asymptotic limits of the degree of electronegativity 
\cite{Lic94,Lic97,Lic00} and added with an ansatz \cite{Kim06}. The ansatz is 
later modified for a better accuracy \cite{Mon08} and one region parabolic 
profile is often neglected \cite{Tho2010,Kem14,Ton15,Kem16}, since the 
flattening of the core becomes more important with the pressure \cite{Kim06}.
Axial edge-to-centre ratios are recently benchmarked against spatially-resolved 
models in argon \cite{Laf15}, and proposed in a new form for electropositive \cite{Rai08} 
and electronegative discharges \cite{Cha16}. Beside the ion losses, neutral wall flux for 
reflective wall might play a very important role on the plasma \cite{Lee95,Kem14}.
The neutral wall flux is either estimated for various geometries \cite{Cha87} 
or calculated by a simpler derivation \cite{Sto95,Kim06,Ton15}. Although, some 
ratios can be partially adaptable for the surface-wave discharges, 
they are not directly applicable in the considered pressure regime due to: (1) the 
structural electronegativity difference 
\cite{Fer88,Dan89,Dan90,Fra99,Fra01}, (2) the detachment dominated 
negative ion loss mechanism of oxygen \cite{Fra02}, 
(3) the unique axial characteristics of the surface-wave discharges 
and (4) the coaxial geometry. 

In this study, we consider the stationary and pulse-modulated surface-wave 
discharges of cylindrical and coaxial structures for a feeding gas of either 
argon or oxygen. We analytically estimate ion and neutral edge-to-centre 
ratios in the considered pressure regime. The computational results are compared 
against a variety of measurements and a good agreement is obtained. The considered 
setups are given in section \ref{sec:set} and the physical model and the 
analytical derivations are explained in section \ref{sec:phy}. The results and 
conclusions are given in sections \ref{sec:res} and \ref{sec:con}, respectively.

%% file: setup.tex
\section{Setup}
\label{sec:set}

\begin{widetext}
\onecolumngrid
\begin{center}
\begin{table}[h!]\footnotesize
\begin{tabular}{lccc|cc}
\\[\dimexpr-\normalbaselineskip+3pt]
\hline
\\[\dimexpr-\normalbaselineskip+3pt]
\\[\dimexpr-\normalbaselineskip+6pt]
{\small  }                &  (I) &  (II) &  (III) & (IV)  & (V)        \\ 
\\[\dimexpr-\normalbaselineskip+6pt]
{\small }               & {Cylindrical}  &    & & {Coaxial}         \\ 
\\[\dimexpr-\normalbaselineskip+6pt]
\hline
\\[\dimexpr-\normalbaselineskip+6pt]
{Type}                &           Surfatron &    Surfatron   &     SLAN  & Plasmaline &     Plasmaline     \\ 
\\[\dimexpr-\normalbaselineskip+6pt]
{Gas}                &           Ar         &   O$_2$                 & O$_2$  & Ar  & Ar or O$_2$      \\ 
$R$ (m)              & $3.0 \times 10^{-3}$& $\; 8.0 \times 10^{-3}  $   & $8.0 \times 10^{-2}$ & $35.0 \times 10^{-3}$ &  $43.0 \times 10^{-3}$ \\ 
\\[\dimexpr-\normalbaselineskip+6pt]
$R_p$ (m)            &                                                   &  & & $15.0 \times 10^{-3}$ &  $6.0 \times 10^{-3}$ \\ 
\\[\dimexpr-\normalbaselineskip+6pt]
$L$ (m)              & $33 \times 10^{-2}$ & $\;  20,34 \times 10^{-2} {\text{ }}^{***}$  & $49.0 \times 10^{-2}$ & $40.0 \times 10^{-2}$ &  $30.7 \times 10^{-2}$ \\ 
\\[\dimexpr-\normalbaselineskip+6pt]
$P$ (Pa)             & $2000$              & $62.7,133.3$                & $50$ & $800$ &  $25-30$ \\ 
\\[\dimexpr-\normalbaselineskip+6pt]
$\cal P$ (W)         & $45$                & $40-107$                    & $2000$ & $400$ &  $300-1500$ \\ 
\\[\dimexpr-\normalbaselineskip+6pt]
$Q$ (sccm)           & $50$                & $32^*$                      & $50$ & $0$ &  $100-200$ \\ 
\\[\dimexpr-\normalbaselineskip+6pt]
$T_h$ (K)            & $800$               & $300$                       & $375$ & $700$ &  $500-900$ \\ 
\\[\dimexpr-\normalbaselineskip+6pt]
$\beta$              & $1.00$              & $1.00$                      & $1.00$ & $0.45$ &  $0.30-0.15^{**}$ \\ 
\\[\dimexpr-\normalbaselineskip+6pt]
{$\cal P$ opr.}      & {\em con.} & {\em con.} & {\em pul.} & {\em con.} & {\em con.} or {\em pul.}       \\ 
\\[\dimexpr-\normalbaselineskip+6pt]
{meas.}              & { RyS, TS}  & { MI, A} & { TALIF} & { TS } & { LP} or { OES}       \\ 
\\[\dimexpr-\normalbaselineskip+3pt]                                                                                          
{Ref. }                &    \cite{Hub12}   &   \cite{Gra89}   &    \cite{Geo02}    & \cite{Rah14_1}  & \cite{Ste13}       \\ 
\hline
\end{tabular}
\caption{The simulation parameters of the considered surface-wave discharges, 
where a Roman number corresponds to each setup, $R$ denotes the outer plasma 
radius, $R_p$ the inner plasma radius of the coaxial structure, $L$ the length 
of the plasma, $P$ the pressure, $\cal P$ the input power, $Q$ the net flow 
rate, $T_h$ the gas temperature, and $\beta$ denotes the power transfer 
efficiency, i.e., the ratio of the absorbed to the input power. 
The operation of the power input is either continuous ({\em con.}) or 
pulse-modulated ({\em pul.}).
The references use a variety of distinct measurement methods (denoted by ``meas.''). 
They are Rayleigh 
Scattering (RyS), Thomson Scattering (TS), Microwave Interferometry (MI), 
Absorption methods (A), time-resolved Two-photon Allowed Laser-Induced 
Fluorescence (TALIF), Langmuir Probe (LP) and Optical Emission Spectroscopy 
(OES).
The parameters are mostly obtained from the corresponding reference. \\
$^*$ Adapted from \cite{Kem15}. \\
$^{**}$ It is estimated with a lower value than the calculation by 
\cite{Rah14_1} due to the structural differences. \\
$^{***}$ We assume that the plasma length does not drastically change for the 
considered pressures.
} 
\label{tab:sets}
\end{table}
\end{center}
\end{widetext}
\twocolumngrid

A series of surface-wave discharges of either a cylindrical or a coaxial structure are 
considered. The plasma dimensions, operation parameters as well as the feeding 
gas of each setup are given in table \ref{tab:sets}. A Roman number is assigned 
to each setup that is used for reference throughout this study. The outer radius 
of the both structures is denoted by $R$ and interior radius in the case of 
the coaxial structure is denoted by $R_p$. The axial plasma dimension is 
represented by $L$ and the plasma volume is represented by $V$. The feeding gas 
is either argon or oxygen and each setup differs in dimension as well as the 
operation parameters.

The operation parameters cover a pressure range of $25-8000\:$Pa and an input 
power range of $45-2000\:$W with either continuous or pulse-modulated power 
input. The discharges are in the collisional regime 
$\lambda_i < (L,2R)$ \cite{Fra99}, where $\lambda_i$ is the ion mean free path, 
or in an alternative stricter form $\lambda_i < \sim (L/2,R) (T_i/T_e)$, which
is assigned to the high-pressure regime of the conventional global model 
implementation \cite{LieBook2005}. The discharges are axially large, $L \gg R$, 
and we frequently use this fact in the simplifying assumptions of the physical 
model.

The cylindrical and coaxial structures have different power transfer 
efficiencies. In the former structure, the power input is completely absorbed by 
the plasma, based on the earlier spatially resolved models \cite{Jim12,Kem15}. 
In the case of the coaxial structure, the efficiency significantly decreases, 
inversely proportional to the power input. We use the calculated efficiency in 
simulating setup (IV)  \cite{Rah14_1} and assume a slightly smaller value for 
setup (V) due to the structural differences. 
The inner radial wall materials of the coaxial structures are quartz.
The coaxial setup 
IV does not have a solid outer boundary, whereas a metal substrate holder forms 
the outer boundary in setup V. The wall recombination probabilities of the quartz 
wall (table \ref{tab:wall}) are assumed at these boundaries due to lack of data, 
after a proper sensitivity analysis (see section \ref{sec:res}).

A particular focus of interest is on the coaxial structure (V) described in 
table \ref{tab:sets}. The discharge is commercially used in the packaging 
industry for the deposition and sterilization purposes \cite{Ste13}. The energy 
coupling is provided by the microwaves, while a radio-frequency substrate bias 
simultaneously controls the ion energy impinging on the surface. The bias 
voltage is switched-off while performing the measurements of the current study
in order to eliminate the additional degree of freedom in defining the power
transfer efficiency. Compared to the conventional plasmalines, the setup 
slightly differs in the radial structure. The outer plasma radius gradually 
increases moving along the surface-wave propagation direction and settles to a 
constant value for the large part of the axial discharge dimension. We assume 
the constant value of the radius in the simulations.

%% file: phys_mod.tex
\section{The physical model} 
\label{sec:phy}

A global model \cite{ChabertBook} is used within the multi-fluid, 
two-temperature plasma description \cite{Bit04}. The model assumes an approximate 
spatial homogeneity inside the plasma volume and converts the 
spatially-resolved description to a zero-dimensional formalism with a 
significant reduction of the numerical load. The chemical kinetics defines the 
local interactions between the plasma species while the wall flux rates 
determine the transport and the plasma-wall interactions. Conventionally, the 
wall flux rates are heuristically estimated from the asymptotic analytical 
solutions of simplified one-dimensional description at comparatively much lower 
pressures \cite{Lic00,Kim06}. Following a similar approach, a set of necessary 
wall flux rates are analytically estimated for the positive ions and the neutral 
particles in the considered pressure regime. The negative ion wall flux is 
assumed to vanish due to the sheath potential drop. In the following, the wall 
flux rates are initially derived based on a one-dimensional simplified 
description and later the global model is described. The derivations are often 
compared to their analogues used in the low-pressure radio-frequency plasma.
For the sake of comparison, an edge-to-centre formalism is adapted, in contrast
to an earlier study \cite{Kem15}.

\subsection{Derivation of the wall flux rates}

The validity of the collisionless models \cite{God86,Kim06} is breached in the 
considered collisionality regime and a constant diffusion model is adopted 
instead \cite{LieBook2005,Rai07,Fer88,Phe90,Lic00}. The one-dimensional model 
is based on the drift-diffusion formulation, where the drift is due to the 
ambipolar electric field, and the ambipolar diffusion simply reduces to the 
regular diffusion for the neutrals. The axial gradient and the axial wall flux 
are neglected due to the large length to radius ratio of the surface-wave 
discharges $L \gg R$. Then, on a reference frame with a vanishing net mass flow 
velocity, steady-state particle continuity equation takes the form
\begin{equation}
 \frac{1}{r} \frac{d}{dr} \left( r D_i \frac{d n_i}{d r}\right) + S_i = 0,
\end{equation}
where $r$ is the radial position, $i$ denotes either a neutral with a reflective 
wall or a positive ion, $n_i$ is particle density,  $D_i$ is the diffusion 
coefficient and $S_i$ is the source. 

\subsubsection{Wall flux estimation for positive ions}

The ambipolar diffusion is defined by the multiple-ion Fick like diffusion 
\cite{Har03,Kem14}, which can be equivalently derived by the assumption of 
Boltzmann equilibrium electrons \cite{Lic00,Bog01}. Together with the assumption 
${\bf \nabla} n^+_i/n^+_i \approx {\bf \nabla} n_e/n_e$ 
($n_e$ and $n^+_i$ are the electron and positive ion densities, respectively)
\cite{Kem15,Fra99} the ambipolar diffusion of the positive ion $i$ takes the 
form
\begin{equation}
D_{i+} = D_i \left( 1 + T_e/T_i \right),
\end{equation}
where $T_e$ is the electron temperature and $T_i$ is the ion temperature. Then
assuming an approximate thermal homogeneity and using the quasineutrality constraint, 
the particle continuity equation of positive ion $i$ is
\begin{equation}
\frac{d^2 n^+_i}{dr^2} + \frac{1}{r} \frac{d n^+_i}{d r} + \frac{\nu^+_i}{D_{i+}(1+\alpha)} n^+_i = 0,
\label{eqn:ion_pb}
\end{equation}
where $\nu^+_i$ is the ionisation frequency and $\alpha$ is the degree of 
electronegativity. Here the ion recombination is neglected and 
$\alpha=0$ denotes the electropositive plasma. 

In this form, an estimation is required for the spatial profile of the degree of 
electronegativity that can be generally characterized by the collisionality.
The analyses at lower collisionality - $\lambda_i > (L/2,R) (T_i/T_e)$ - show 
that the value of the central degree of electronegativity $\alpha_0$ defines 
the spatial structure of the discharge. At low values of $\alpha_0$, the plasma is structured with a parabolic 
electronegative core surrounded by an electropositive edge \cite{Lic94}. The 
parabolic core enlarges and flattens with increasing electronegativity 
\cite{Kou96,Lic97,Lic00}. However, in the considered collisional regime 
- $\lambda_i < \sim (L/2,R) (T_i/T_e)$ - a ratio of the attachment to the 
ionization frequency $P=\frac{\mu_+}{\mu_-}\frac{\nu_i^a}{\nu_i^+}$ 
defines the spatial dependence of the degree of electronegativity 
($\mu_+$ and $\mu_-$ are the mobilities of positive and negative ions, 
respectively) \cite{Fer88,Fra99}. In the asymptotic limits of very small and 
very large values of $P$, 
$\alpha$ is spatially 
homogeneous and otherwise piecewise homogeneity is a valid assumption in a good 
approximation \cite{Fer88,Dan89,Dan90,Fra99}. Consequently, we assume that 
$\alpha$ is spatially homogeneous, where the 
piecewise homogeneity does not alter the general solution.

For a homogeneous degree of electronegativity, the general solution of equation 
\ref{eqn:ion_pb} can be written as
\begin{equation}
n^+_i(r) = C_1 J_0 \left( \chi r/R \right) + C_2 Y_0 \left( \chi r/R\right),
\label{eqn:ana_sol}
\end{equation}
where $J_0,Y_0$ are the zeroth order Bessel functions of the first and the 
second kind, $\chi,C_1,C_2$ are the constants to be fixed by the boundary 
conditions as well as the normalization. The solution is subjected to the 
boundary conditions $n^+_i(r=0)\neq \infty$, $n^+_i(r=R)=0$ for cylindrical and 
$n^+_i(r=R)=n^+_i(r=R_p)=0$ for coaxial plasma. The particular solution in 
cylindrical structure is in agreement with the analytic estimations in the 
asymptotic limits of very large \cite{Dan89} and very small values of $P$ 
\cite{Fer88,Dan90}. 

In the cylindrical structure, the ion flux density at the Bohm point can be 
written in the form
\begin{equation} 
\left. \Gamma_i^+  \right|_{r_B} = u_{iB} h^+_{iR} n^+_{i0}, 
\label{eqn:flu_cyl}
\end{equation}
where $r_B$ is the radial Bohm point, $n^+_{i0}$ is the central ion density, 
$h^+_{iR}$ is the edge-to-centre ratio of the cylindrical structure and $u_{iB}$ 
is the electropositive Bohm velocity. An estimation of the edge-to-centre ratio 
is (see Appendix \ref{sec:app} for the derivation.)
\begin{widetext}
\onecolumngrid
\begin{equation}
h^+_{iR} \approx  \frac{1}{1+\alpha} (u_{iBE}/u_{iB}) \left( 1 + \left( \frac{R u_{iBE}}{\chi_{01} D_{i+} J_1(\chi_{01})}\right)^2 \right)^{-1/2}
\label{eqn:ha_cly}
\end{equation}
\end{widetext}
\twocolumngrid
where $\chi_{01}$ is the first root of $J_0$, $u_{iBE}$ is the Bohm velocity 
modified by the electronegativity - see section \ref{sec:Bohm} - and 
${1}/{(1+\alpha)}$ is the normalization in the electronegative discharges 
\cite{Lee95}. Equations \ref{eqn:ha_cly} and \ref{eqn:flu_cyl} are valid in the 
electropositive plasma for a vanishing degree of electronegativity.

An analogue of the edge-to-centre ratio that is used in the low-pressure plasma 
models ($\lambda_i > (L/2,R) (T_i/T_e)$) is given by the ansatz
$\bar{h}^+_{iR}=\left( (\bar{h}^+_{iR,a})^2+(\bar{h}^+_{iR,c})^2 \right)^{1/2}$ 
\cite{Tho2010,Kem16} together with
\begin{equation}
\begin{array}{l l}
\bar{h}^+_{iR,a} & \approx \frac{0.8}{1+\alpha_0} \left( 4 + \frac{R}{\lambda_i} + \left( \frac{0.8 R u_{iB}}{\chi_{01} J_1(\chi_{01}) D_{i+}} \right)^2 \right)^{-1/2},  \\
\bar{h}^+_{iR,c} & \approx  \left( \frac{T_e}{T_h} + \left(\frac{T_e}{T_h}\right)^{1/2}  \left({\frac{15}{56} \frac{v_{i+}}{k_{ir,+} \lambda_i}}\right)^{1/2} n_{i+} n_{-}^{-3/2} \right)^{-1}, 
\label{eqn:ha_cly_pre}
\end{array}
\end{equation} 
where a bar notation is used to discriminate from equation \ref{eqn:ha_cly}, 
$n_-$ is the negative ion density, $v_{i+}$ is the mean thermal velocity, 
$k_{ir,+}$ is the rate coefficient of ion-ion recombination and $\lambda_i$ is 
the ion mean free path. The first term of the ansatz $\bar{h}^+_{iR,a}$ denotes 
a parabolic electronegative core surrounded by an electropositive edge and it is 
valid at a low degree of electronegativity for a large range of collisionality 
regimes \cite{LieBook2005}. The second term $\bar{h}^+_{iR,c}$ denotes a single 
electronegative region of a heuristic flat-topped description and it is valid at 
a large degree of electronegativity for recombination dominated negative ion 
loss \cite{Fra02}.

The edge-to-centre ratio $\bar{h}^+_{iR,a}$ is equivalent to $h^+_{iR}$ in the 
considered collisional regime. The ratio $\bar{h}^+_{iR,a}$ is also 
alternatively applicable in this regime since it covers a large range of 
collisionality and the lower asymptotic limits of both $P$ and $\alpha_0$ are 
substitutes for each other \cite{Fer88}. A non-structured highly electronegative 
plasma that is heuristically expressed by $\bar{h}^+_{iR,c}$ \cite{Lic97} is 
already contained in $h^+_{iR}$ with the assumption of spatially invariant 
degree of electronegativity at a large value of $P$ in the collisional regime 
\cite{Fra99}. Additionally, a non-structured region does not necessarily require 
a large degree of electronegativity in a collisional plasma \cite{Fer88,Fra99} 
as imposed by $\bar{h}^+_{iR,c}$ and considered discharges are not strongly 
electronegative with a maximum value of $\alpha=12$. As a result, the ratio 
$\bar{h}^+_{iR,c}$ is not strictly adapted in a similar ansatz in the 
collisional regime. A numerical comparison of $h^+_{iR}$ and $\bar{h}^+_{iR}$ is
given in section \ref{sec:res} for the considered cylindrical plasma sources.

The coaxial structure has multiple Bohm points assigned to each radial boundary 
and the positive ion flux densities at these Bohm points can be written as 
\begin{equation} 
\begin{array}{l l}
\left. \Gamma_i^+  \right|_{r_B} & = u_{B}  h^+_{iR} n^+_{i0}, \\
\\[\dimexpr-\normalbaselineskip+8pt]
\left. \Gamma_i^+  \right|_{r_{B_p}} & = - u_{B}  h^+_{iR_p} n^+_{i0} 
\end{array}
\end{equation} 
where $r_B$ denotes the Bohm point near the outer boundary at $R$, $r_{B_p}$ is 
the Bohm point near the inner boundary at $R_p$ and the negative sign at 
$r_{B_p}$ is due to the direction of the radial unit vector. Similar to the 
cylindrical structure, the radial edge-to-centre ratio of each radial boundary 
is derived as (see Appendix \ref{sec:app} for the derivation.)
\begin{widetext}
\onecolumngrid
\begin{equation}
\begin{array}{l l}
h^+_{iR} & \approx \frac{1}{1+\alpha} (u_{iBE}/u_{iB}) \left(  1 + \left( \frac{R u_{iBE}}{\chi D_{i+} (C_1 J_1(\chi) + C_2 Y_1(\chi))}\right)^2 \right)^{-1/2}, \\
\\[\dimexpr-\normalbaselineskip+8pt]
h^+_{iR_p} & \approx  \frac{1}{1+\alpha} (u_{iBE}/u_{iB}) \left(  1 + \left( \frac{R u_{iBE}}{\chi D_{i+} (C_1 J_1(\chi R_p/R) + C_2 Y_1(\chi R_p/R))}\right)^2 \right)^{-1/2}, 
\label{eqn:ha_coa}
\end{array}
\end{equation}
\end{widetext}
\twocolumngrid
where $\chi,C_1$ and $C_2$ are defined by the normalization and the boundary 
conditions. 

\subsubsection{Bohm velocity in an electronegative plasma} 
\label{sec:Bohm}
A new expression is required for the Bohm velocity in the presence of the 
negative ions. In order to calculate this expression, it is commonly assumed 
that the negative ions, denoted by $n^-_i$, are in Boltzmann equilibrium 
\cite{Tho59} at low-pressure 
\begin{equation}
T_i {\bf \nabla} n^-_i/n^-_i = T_e {\bf \nabla} n_e/n_e. 
\end{equation}
This assumption introduces an additional factor to the Bohm velocity 
\cite{Bra88,LieBook2005} 
\begin{equation}
\left( \frac{1+\alpha_s}{1+\alpha_s \gamma} \right)^{1/2},
\end{equation}
where $\gamma$ is the ratio of electron to the negative ion temperatures and
$\alpha_s$ is the degree of electronegativity at the sheath.

The validity of the Boltzmann equilibrium expires at high-pressure 
\cite{Lic00,Fra01,Cha16} and an alternative assumption is valid 
\cite{Phe90,Fra99}
\begin{equation}
{\bf \nabla} n^-_i/n^-_i = {\bf \nabla} n_e/n_e.
\end{equation}
This assumption completely eliminates the additional factor on the Bohm velocity 
by simply equating $\gamma$ to unity. The transition between these two 
assumptions can be defined by the condition of attachment dominance \cite{Bog01}
\begin{equation}
\tau_{an} K_a > 1 
\end{equation}
where $\tau_{an}$ is the time-scale of the ambipolar diffusion and $K_{a}$ is the 
electron attachment frequency. The following form for the Bohm velocity is used 
based on this condition
\begin{equation}
u_{iBE}/u_{iB} = 
\begin{cases}
\left( \frac{1+\alpha}{1+\alpha \gamma} \right)^{1/2} & \text{if } \tau_{an} K_a \leq 1, \\
1  & \text{if } \tau_{an} K_a > 1, 
\end{cases}
\end{equation}
where the assumption of the spatially homogeneous degree of electronegativity is 
implemented. A comparison of these expressions on the computational results is 
discussed in section \ref{sec:res}.

\subsubsection{Wall flux estimation for neutrals}
The axial gradients and the axial wall flux are neglected due to the large 
length to the radius ratio. Assuming that the neutral source $S_i$ is radially 
homogeneous as an approximation \cite{Kim06,Sto95}, the stationary neutral 
continuity equation can be written as
\begin{equation}
\frac{1}{r} \frac{d}{dr} \left( r \frac{d n^N_i}{d r}\right) + \frac{S_i}{D_i} = 0,
\end{equation}
where $n^N_i$ is the neutral density with a reactive wall, $S_i$ is the source 
and $D_i$ is the diffusion and superscript $^N$ represents the neutral particles.
The general solution can be written as
\begin{equation}
n^N_i(r) = C \left( 1- \frac{r^2}{a^2} + b \ln(r) \right),
\end{equation}
where $C, a$ and $b$ are the constants defined by the normalization and the 
boundary conditions. The flux boundary conditions are set on the chamber walls 
based on the reactivity of the wall material and a finite solution is 
imposed at the radial centre of the cylindrical structure.

The flux boundary condition at the radial wall of a cylindrical structure is 
\cite{Cha87}
\begin{equation}
 \left. \left( -D_i \frac{d n^N_i}{d r}\right)\right|_{r=R}= D_i \left. n^N_i\right|_{r=R} \frac{1}{\l_{iR}},
\end{equation}
where $\l_{iR}$ is the linear extrapolation length assigned to the wall. The 
linear extrapolation length can be written as
\begin{equation}
{\l_{iR}}=\frac{2 D_i}{v_i}\frac{2-\gamma_{iR}}{\gamma_{iR}},
\end{equation}
where $\gamma_{iR}$ is the wall reaction probability and $v_i$ is the thermal 
velocity. The boundary conditions of a cylindrical structure set $b=0$ and
$a^2=R^2 + 2 R \l_{iR}$ and reduce the neutral wall flux density to the form
\begin{equation}
 \left. \Gamma_i^N  \right|_{R}= \frac{D_i}{\l_{iR}}  h^N_{iR}   n_{i0},
\label{eqn:neu_flu_cyl}
\end{equation}
where $n_{i0}$ is the density at the radial centre and neutral edge-to-centre 
ratio is 
\begin{equation}
h^N_{iR}=\left( 1+\frac{R}{2\l_{iR}} \right)^{-1}.
\label{eqn:neu_hra_cyl}
\end{equation}

An expression is similarly derived by Kim {\em et al} \cite{Kim06} for radially large 
cylindrical plasma chambers, $R \gg L$. Although, the neutral continuity 
equation is governed axially for such a chamber, the derived edge-to-centre 
ratio is equal to $h^N_{iR}$ when the discharge radius $R$ and the length $L$ 
are interchanged. A net wall flux rate is also estimated by Chantry \cite{Cha87} 
ignoring the source term of the continuity equation. This rate can be written in 
the form of equation \ref{eqn:neu_flu_cyl} with the following edge-to-centre 
ratio 
\begin{equation}
\bar{h}^N_{iR}=\left( 1+ \frac{ \Lambda_0^2}{\l_{iR}} \frac{A}{V} \right)^{-1}, 
\end{equation}
where $\Lambda_0$ is the effective diffusion length and a bar notation is used 
to distinguish from equation \ref{eqn:neu_hra_cyl}. For an axially large 
cylindrical discharge, $L \gg R$; this ratio can be approximated by 
$
\bar{h}^N_{iR} \approx \left( 1+ \left( \frac{ R}{ 2 \l_{iR}} \right) \left( \frac{2}{2.405} \right)^2 \right)^{-1} 
$
that differs from $h^N_{iR}$ only with a factor of $(2/2.405)^2$ in the second 
term. A further numerical comparison of the considered plasma sources is 
provided in section \ref{sec:res}.

The wall flux densities of the coaxial structure on the radial walls $R$ and 
$R_p$ can be similarly written in the form
\begin{equation}
\begin{array}{l l}
 \left. \Gamma_i^N  \right|_{R}  & = \frac{D_i}{\l_{iR}} h^N_{iR} n^N_{i0}, \\
\\[\dimexpr-\normalbaselineskip+8pt]
  \left. \Gamma_i^N  \right|_{R_p}  & = -  \frac{D_i}{\l_{iR_p}} h^N_{iR_p}  n^N_{i0},
\end{array}
\end{equation} 
where the linear extrapolation length generally differs between the inner and 
outer walls due to distinct reaction probabilities of the wall materials. The 
corresponding boundary conditions set non-zero values for the constants $b$ and 
$a$ that depends on the positions and the extrapolation lengths of the 
walls. Accordingly, the inner and outer edge-to-centre ratios with a 
normalization constant $C$ are
\begin{widetext}
\onecolumngrid
\begin{equation}
\begin{array}{l l}
h^N_{iR} & = 
C \frac{\l_{iR} \left((R_p-2 \l_{iR_p}) (R_p-R) (R_p+R)-2 R_p R^2 \ln \left(\frac{R_p}{R}\right)\right)}{\l_{iR} R_p^2 (R_p-2 \l_{iR_p})-R_p R^2 (2 \l_{iR}+R) \ln \left( \frac{R_p}{R} \right) +\l_{iR_p} R^3+2 \l_{iR_p} \l_{iR} R^2} \\
\\[\dimexpr-\normalbaselineskip+8pt]
h^N_{iR_p} & = 
C \frac{\l_{iR_p} \left( 2 R_p^2 R \ln \left(\frac{R_p}{R}\right)-(R_p-R) (R_p+R) (2 \l_{iR}+R) \right) }{\l_{iR} R_p^2 (R_p-2 \l_{iR_p})-R_p R^2 (2 \l_{iR}+R) \ln \left( \frac{R_p}{R} \right) +\l_{iR_p} R^3+2 \l_{iR_p} \l_{iR} R^2}.
\label{eqn:neu_flu_coa}
\end{array}
\end{equation} 
\end{widetext}
\twocolumngrid
The ratio at each wall depends on the positions and the reactive properties of 
the both walls as well as the transport properties of the species $i$.

%% file: globalmodel.tex
\subsection{The global model}
\label{subsec:gm}

The plasma is described by the volume-averaged particle and electron energy 
continuity equations with an assumption of Maxwellian electron energy 
distribution function. The plasma sheath is ignored in the model and it is 
considered only in estimating energy loss due to ion wall flux \cite{LieBook2005}.
The gas temperature is externally provided and a unit convention of the 
two-temperature description is adopted from an earlier study \cite{Kem16}
unless stated otherwise. 

\subsubsection{Volume-averaged particle continuity equation}
The volume-averaged particle continuity equation is written in the form
\begin{equation}
\frac{d N_i}{d t} = \left. \sum_{j}  {\cal W}_{ij}R^{j}_{i} \right|_{V} 
+ \left. \sum_{j}  {\cal W}_{ij}R^{j}_{i} \right|_{W},
\end{equation}
where $i$ denotes the particular species, $N_i$ is the volume-averaged particle 
density, $j$ denotes a particular source channel with a rate $R^j_i$ and a net 
stoichiometric coefficient ${\cal W}_{ij}$. The subscript ``$_V$'' represents 
the chemical reactions inside the plasma volume and ``$_W$'' represents the wall 
losses due to the net mass flow as well as the net ion and neutral wall flux.

The sets of species - Ar or O$_2$ - and the chemical reactions are adapted from 
earlier studies \cite{Gud07,Jim12,Kab07,Kem15} and they are tabulated in 
\ref{sec:app2}. The flow-in rate of the feeding gas - either Ar or O$_2$ in the 
current study - due to the net mass flow rate $Q$ (sccm) is
$
\left. R_{\text{Ar,O}_2}^{\text{F-in}} \right|_{W} =  c \frac{QP_{\text{atm}}}{V \text{k}_{B} T_{\text{in}}}, 
$
where $c=1.6667 \times 10^{-8} \: \: {\text{sccm}}/{\text{m}^3 \text{s}^{-1}}$ 
is the conversion factor $P_{\text{atm}}$ is the atmospheric pressure and 
$T_{\text{in}}=300 \: $K is the input gas temperature. The mass flow imposes a 
flow-out rate for a particle $i$
$
\left. R_i^{\text{F-out}} \right|_{W} = c  Q \frac{P_{\text{atm}}T_h}{V P  T_{\text{in}}} n_i.
$

The net wall loss rates follow the earlier derivation of the 
wall flux. The wall loss rates of the ions $R^+_i \left. \right|_W$ and of the 
neutrals $R^N_i \left. \right|_W $ in the cylindrical structure are written as
\begin{equation} 
\begin{array}{l l}
R^+_i \left. \right|_W & = u_{iB}  \frac{h^+_{iR} A_R}{V}  N_{i}, \\
\\[\dimexpr-\normalbaselineskip+4pt]
R^N_i \left. \right|_W & = \frac{D_n}{\l_{iR}} \frac{h^N_{iR} A_R}{V} N_{i},
\end{array} 
\label{eqn:cyl_wal}
\end{equation} 
where $A_R$ is the radial wall area and the edge-to-centre ratios - $h^+_{iR}$,  
$h^N_{iR}$ - are given by equations \ref{eqn:ha_cly} and \ref{eqn:neu_hra_cyl}, 
respectively. The loss rates at the outer (at $R$) and the inner (at $R_p$) 
walls of the coaxial structure take the form
\begin{equation} 
\begin{array}{l l}
 R^+_i \left. \right|_W & =  \left( u_{iB} \frac{h^+_{iR} A_R}{V} + u_{iB} \frac{h^+_{iR_p}A_{R_P}}{V} \right) N_{i}, \\
 \\[\dimexpr-\normalbaselineskip+4pt]
 R^N_i \left. \right|_W & = \left( \frac{D_i}{\l_{iR}} \frac{h^N_{iR} A_R}{V} + \frac{D_i}{\l_{iR_p}} \frac{h^N_{iR_p} A_{R_p}}{V} \right)  N_{i},
 \end{array}
\label{eqn:coa_wal}
\end{equation} 
where $A_R$ is the outer wall surface area, $A_{R_p}$ is the inner wall surface 
area and the edge-to-centre ratios - $h^+_{iR}$,  $h^+_{i{R_p}}$ and 
$h^N_{iR}$, $h^N_{i{R_p}}$ - are given in equations \ref{eqn:ha_coa} and 
\ref{eqn:neu_flu_coa}, respectively.

\subsubsection{Volume-averaged electron energy continuity equation}
The volume-averaged electron energy continuity equation for a Maxwellian 
electron energy distribution function can be written as 
\begin{equation}
\frac{d}{d t} \left( \frac{3}{2} n_e T_e \right) = Q_{\text{abs}} - (Q_{\text{Ine}} + Q_{\text{Ela}} + Q_W),
\end{equation}
where $Q_{\text{abs}}$ is the power absorbed by electrons and the energy loss 
rates are due to chemical reactions $Q_{\text{Ine}}$, elastic collisions 
$Q_{\text{Ela}}$ as well as net wall flux $Q_W$.

A power transfer efficiency, $\beta$, defines the power absorbed by the 
electrons $Q_{\text{abs}}= \beta {\mathcal P}/V$ for an input power of 
${\mathcal P}$. The inelastic energy loss is 
$Q_{\text{Ine}}= \left. \sum_j {\cal E}_j R^j_e \right|_V$, where $R^j_e$ is an 
electronic reaction rate and ${\cal E}_j$ is the reaction energy calculated from 
the internal energies of the particles \cite{Kem14_2}. The elastic loss is 
$Q_{\text{Ela}} = \sum_{i} n_e n_i 3  T_e \frac{m_e}{m_i} k^{\text{El}}_{ei}$, 
where $k^{\text{El}}_{ei}$ is the elastic rate coefficient calculated from the 
corresponding cross-section. 

The wall energy loss is \cite{LieBook2005}
\begin{equation}
Q_W = 
\left. \sum_{i \in Ions} \left( {\cal E}_P 
+ {\cal E}_e + {\cal E}_s  \right) R^+_{i} \right|_{W},
\end{equation}
where ${\cal E}_P$ is the plasma potential, ${\cal E}_e$ is the mean energy loss
per electron and ${\cal E}_s$ is the sheath potential. The plasma potential is 
\cite{LieBook2005}
\begin{equation}
{\cal E}_P = \frac{1}{2}  \left( u_{+,BE}/u_{+,B}  \right)^2  T_e,
\end{equation} 
where $u_{+,BE}$ is the Bohm velocity of the dominant ion while the mean energy 
loss per electron is ${\cal E}_e=2T_e$. We use an estimation of the sheath 
potential by Thorsteinsson  {\em et al} \cite{Tho2010_2} together with the 
assumption of the spatially homogeneous degree of electronegativity
\begin{equation}
{\cal E}_s \approx \ln \left( 4 \frac{u_{+,BE}}{v_e} \frac{1+ \alpha}{1+\alpha ( {v_{-}}/v_e)^2} \right) T_e,
\label{eqn:esheath}
\end{equation}
where $v_e$ and $v_{-}$ are the mean thermal velocities of electrons and ions, 
respectively. The assumption of the Boltzmann equilibrium of the negative ions - 
or its high-pressure alternative ${\bf \nabla} n^-_i/n^-_i = {\bf \nabla} n_e/n_e$ - does not significantly affect the sheath 
potential estimation except at very large degree of electronegativity 
\cite{Tho2010_2}. The sheath and plasma potential expressions for the 
electropositive gas are obtained for a vanishing degree of electronegativity.

%% file: results.tex
\section{Results}
\label{sec:res}

A variety  of cylindrical and coaxial surface-wave discharges are simulated 
and the simulation results are compared with the available spatially-averaged 
measurements. The spatially-resolved measurements are mostly provided in their 
original form for a detailed view of the profile. The operation parameters of 
the simulated discharge setups are given in table \ref{tab:sets}. The 
measurements are denoted by symbols, where filled symbols are reserved for their
spatial-average, and the calculations are shown by lines or line-points. A 
discussion on the state of certain model parameters and on the analytically 
estimated radial profiles as well as a comparison of the edge-to-centre ratios 
against their low collisionality analogues is also provided at the end of this 
section. 

\begin{figure}[!ht]
\minipage{0.33\textwidth}
  \includegraphics[width=0.9\linewidth]{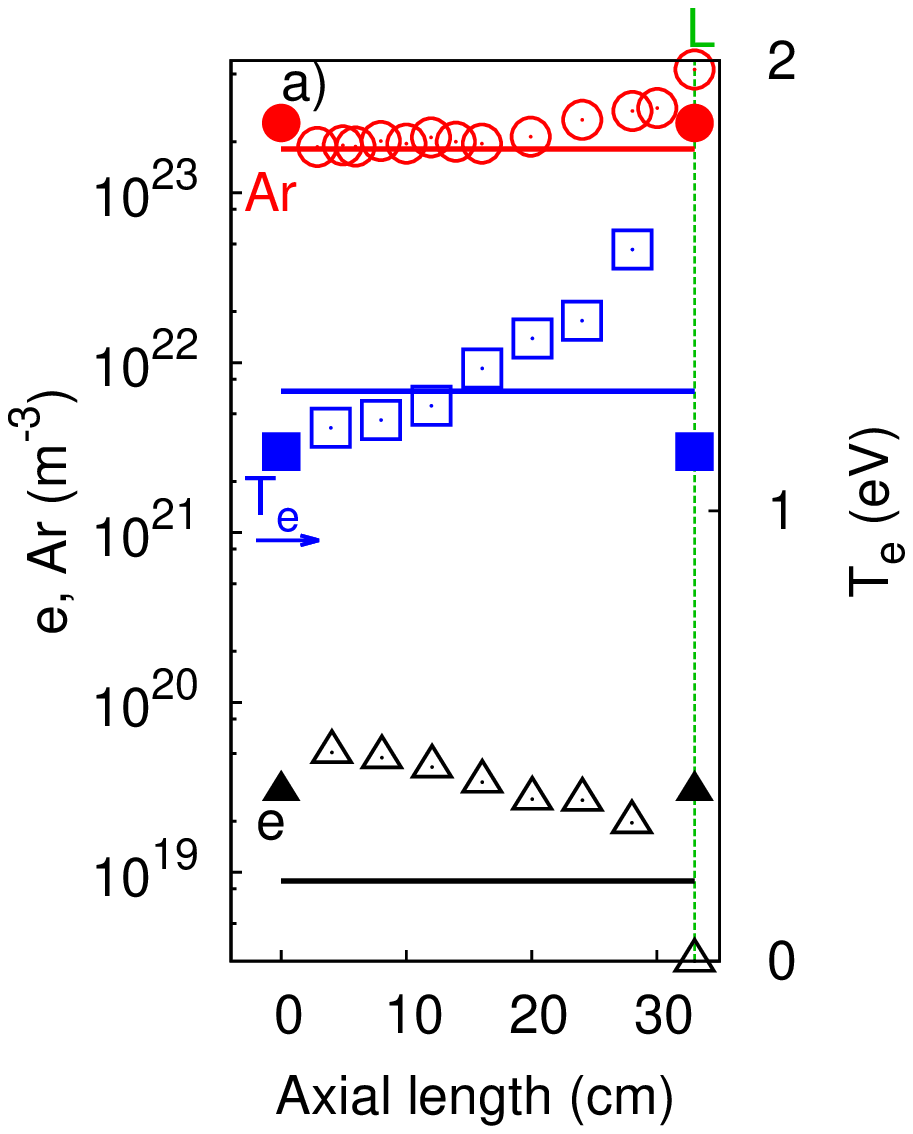}
\endminipage\hfill
\minipage{0.33\textwidth}
  \includegraphics[width=0.9\linewidth]{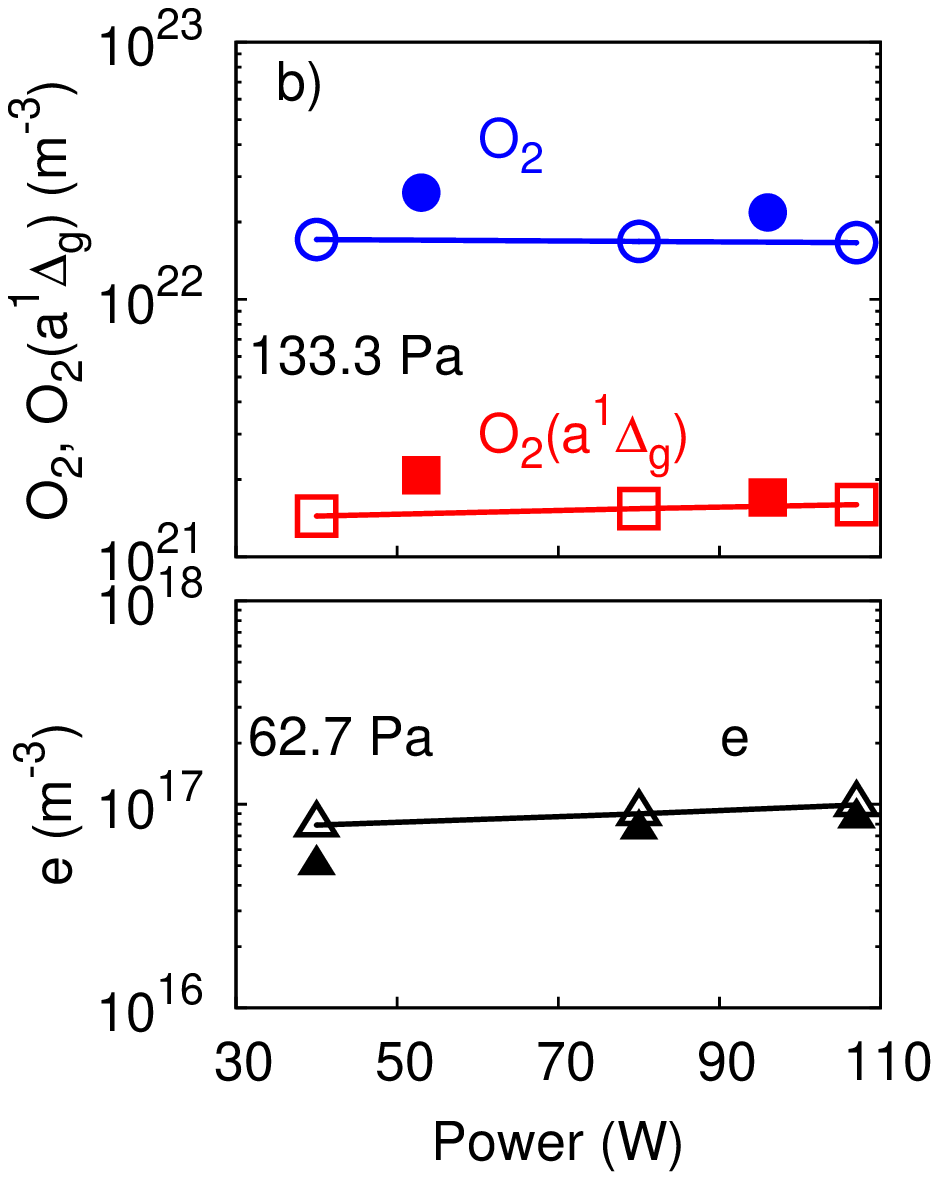}
\endminipage\hfill
\minipage{0.33\textwidth}
  \includegraphics[width=0.9\linewidth]{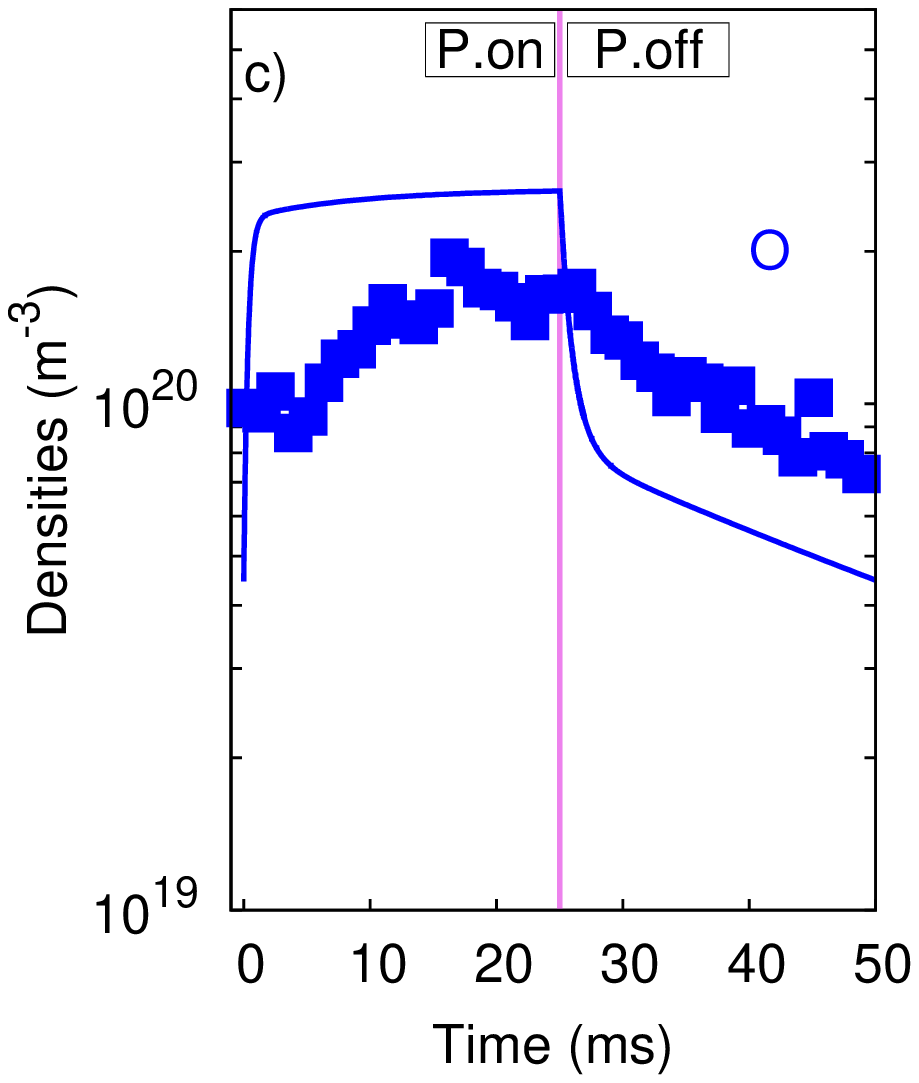}
\endminipage
\caption{The measurements performed on distinct surface-wave discharges of 
cylindrical structures I \cite{Hub12} , II \cite{Gra89} and III \cite{Geo02} of table  \ref{tab:sets} and the simulation 
results (line or line-points). 
The simulation results are volume-averaged and they are comparable to the 
spatially-averaged measurements.
a) Axially-resolved e ($\triangle$) and Ar 
(\textcolor{red}{$\circ$}) densities as well as the electron temperature T$_e$ 
(\textcolor{blue}{$\square$}) in an argon surfatron (I) together with their 
axial averages - denoted by $\blacktriangle$, \textcolor{red}{$\bullet$} and
\textcolor{blue}{$\blacksquare$}, respectively. b) Axial-averages of e 
($\blacktriangle$), O$_2$ (\textcolor{blue}{$\bullet$}) and 
O$_2(\text{a}^1 \Delta_\text{g})$ (\textcolor{red}{$\blacksquare$}) density 
measurements in an oxygen surfatron (II). c) Transient O density measurements 
in a microwave source of SLAN type with a pulse-modulated power input (III). 
}\label{fig:comp_cyl}
\end{figure}

The measurements and the calculations of the cylindrical structure are shown in 
figure \ref{fig:comp_cyl} for a series of distinct discharges fed with either 
argon or oxygen. The operation parameters of each setup are provided in table 
\ref{tab:sets} - under columns I, II, and III. Figure \ref{fig:comp_cyl}-a) 
shows the axially-resolved and axially-averaged electron and 
argon densities as well as the electron temperature in a stationary argon 
surfatron (I) \cite{Hub12}. Figure \ref{fig:comp_cyl}-b) shows axially-averaged e, 
O$_2(a^1 \Delta_g)$ and O$_2$ densities 
of a stationary oxygen surfatron (II) \cite{Gra89} for a set of power input and pressure values.
Figure \ref{fig:comp_cyl}-c) shows transient O density 
on a pulse-modulated microwave source of SLAN type of oxygen (III) \cite{Geo02}. 
We observe a good agreement between the spatially-averaged measurements 
and the simulation results, however, the agreement is only fair for the electron 
density of argon surfatron in figure \ref{fig:comp_cyl}-a). The argon surfatron 
shows radial contraction at the end of the plasma column \cite{Hub13} and this 
might cause such a deviation. The simulation results of the oxygen surfatron 
show a better agreement compared to the earlier results of a rough ion 
edge-to-centre ratio estimation \cite{Kem15}, yet a very similar two-scaled 
decay is observed in the microwave source of SLAN type.

\onecolumngrid
\clearpage

\begin{figure}[!htb]
\minipage{0.32\textwidth}
  \includegraphics[width=0.9\linewidth]{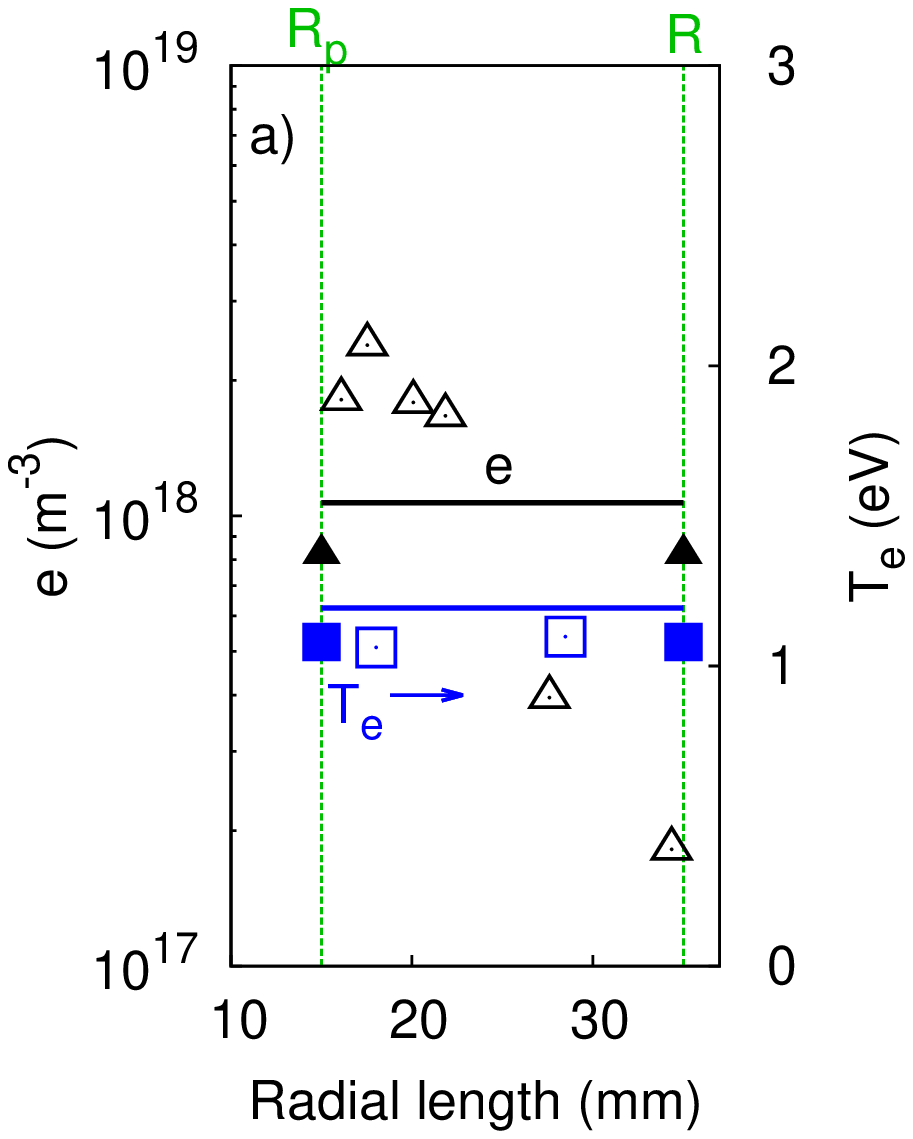}
\endminipage\hfill
\minipage{0.32\textwidth}
  \includegraphics[width=0.9\linewidth]{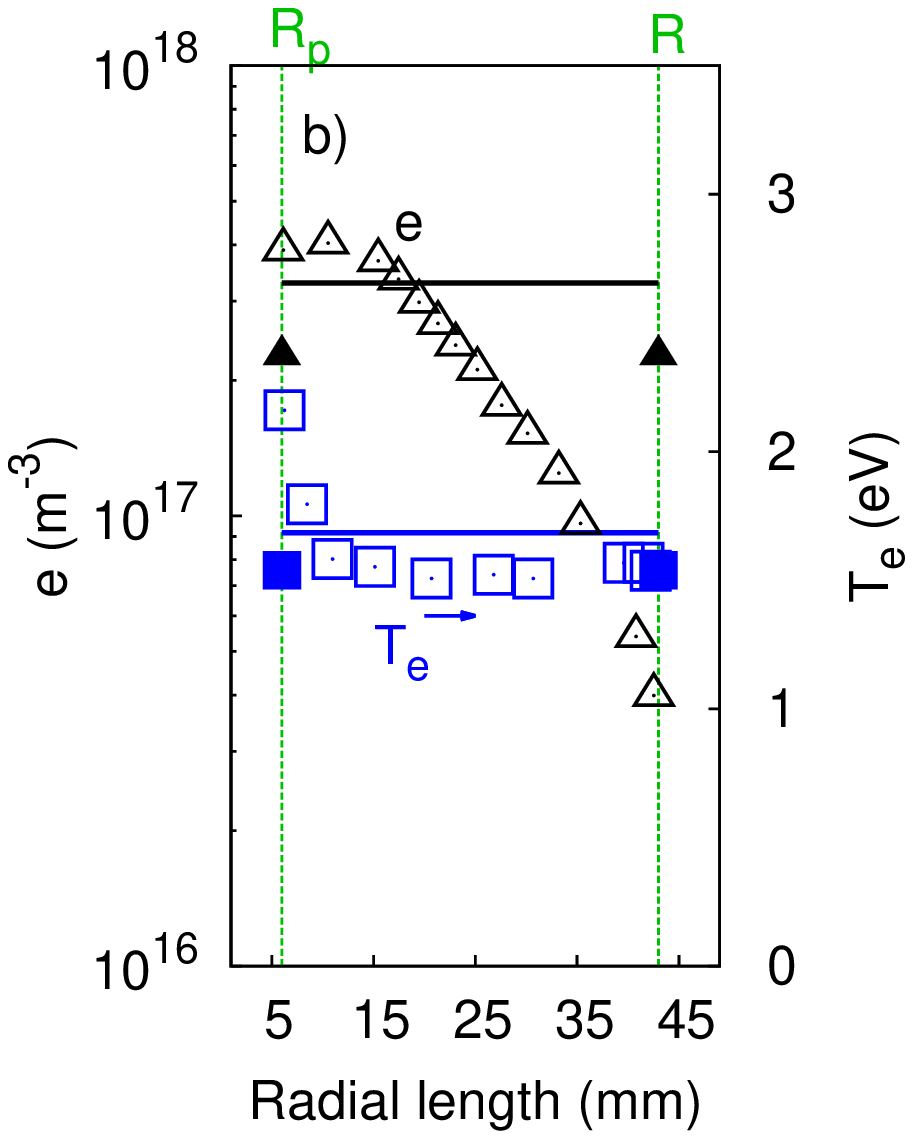}
\endminipage\hfill
\minipage{0.32\textwidth}
  \includegraphics[width=0.9\linewidth]{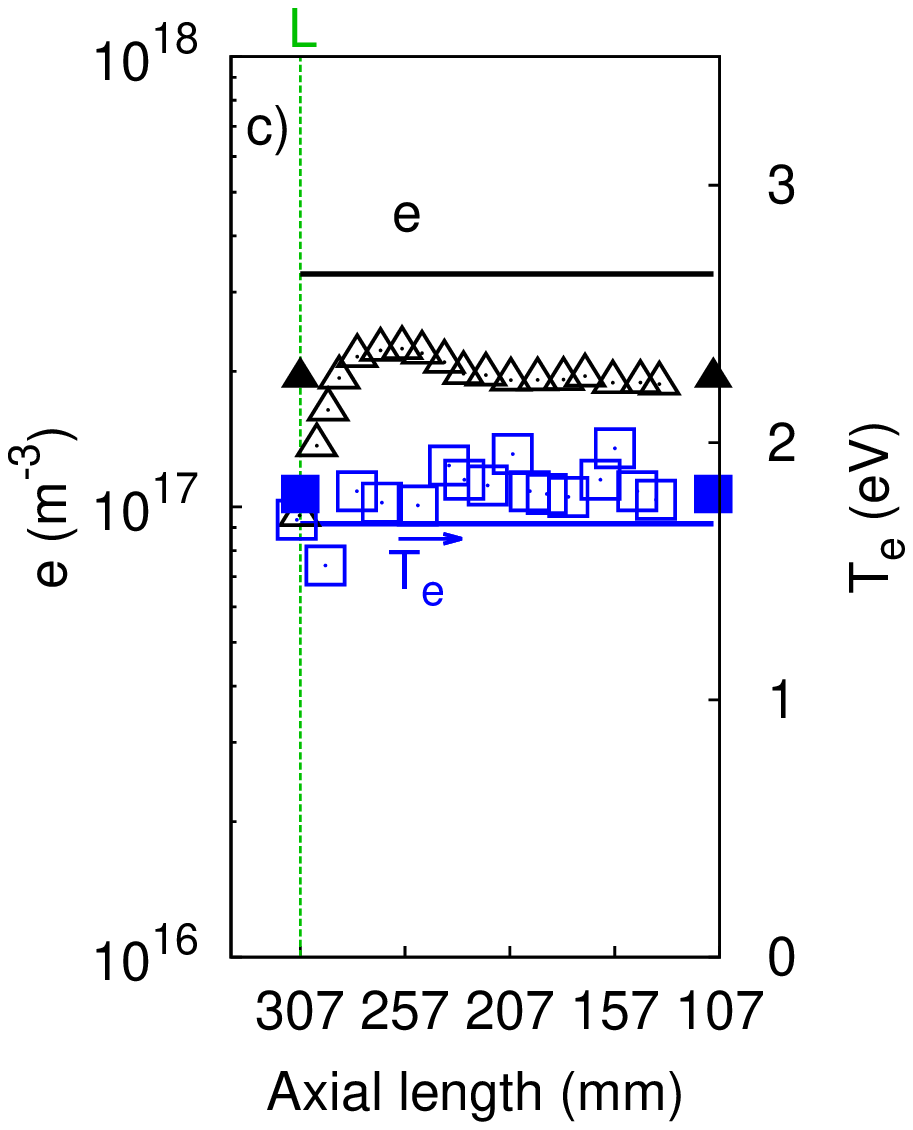}
\endminipage
\caption{The measurements performed on distinct coaxial structures of IV \cite{Rah14_1} and 
V - see table \ref{tab:sets} - and the simulation results (line) for a feeding 
gas of argon. 
The simulation results are volume-averaged and they are comparable to the 
spatially-averaged measurements.
a) Radially-resolved electron density ($\triangle$) and 
temperature (\textcolor{blue}{$\square$}) measurements on the discharge setup IV. 
The radial averages of the electron density and temperature measurements are 
denoted by $\blacktriangle$ and \textcolor{blue}{$\blacksquare$}, respectively.
b) Radially-resolved electron density ($\triangle$) and temperature 
(\textcolor{blue}{$\square$}) measurements as well as their radial averages 
(denoted by $\blacktriangle$ and \textcolor{blue}{$\blacksquare$}, respectively)
on the setup V. c) Axially-resolved electron density ($\triangle$) and 
temperature (\textcolor{blue}{$\square$}) measurements as well as their axial 
averages (denoted by $\blacktriangle$ and \textcolor{blue}{$\blacksquare$}, 
respectively) on the setup V.}\label{fig:comp_coacw}
\end{figure}
\twocolumngrid

The simulation results and the measurements of the coaxial structures IV and 
V - see table \ref{tab:sets} - for a feeding gas of argon are shown in figure 
\ref{fig:comp_coacw}. The radially-resolved and radially-averaged electron 
density and temperature are shown in
figure \ref{fig:comp_coacw}-a) for a stationary plasmaline IV \cite{Rah14_1}.
The same type of data is shown in figure \ref{fig:comp_coacw}-b) in the 
plasmaline of interest V at $30\:$Pa for an input power of $300\:$W with the 
gas temperature of $500\:$K and the gas flow rate of $100\:$sccm, whereas 
axially-resolved and axially-averaged versions are shown in figure 
\ref{fig:comp_coacw}-c). We observe that the radial variation is much more 
amplified than the axial variation and the spatially-resolved measurements are 
in the same order of magnitude with their spatial-averages. The simulation 
results show a good agreement with the stationary radially- and axially-averaged 
measurements. 

\onecolumngrid
\begin{widetext}

\begin{figure}[!htb]
\minipage{0.32\textwidth}
  \includegraphics[width=0.9\linewidth]{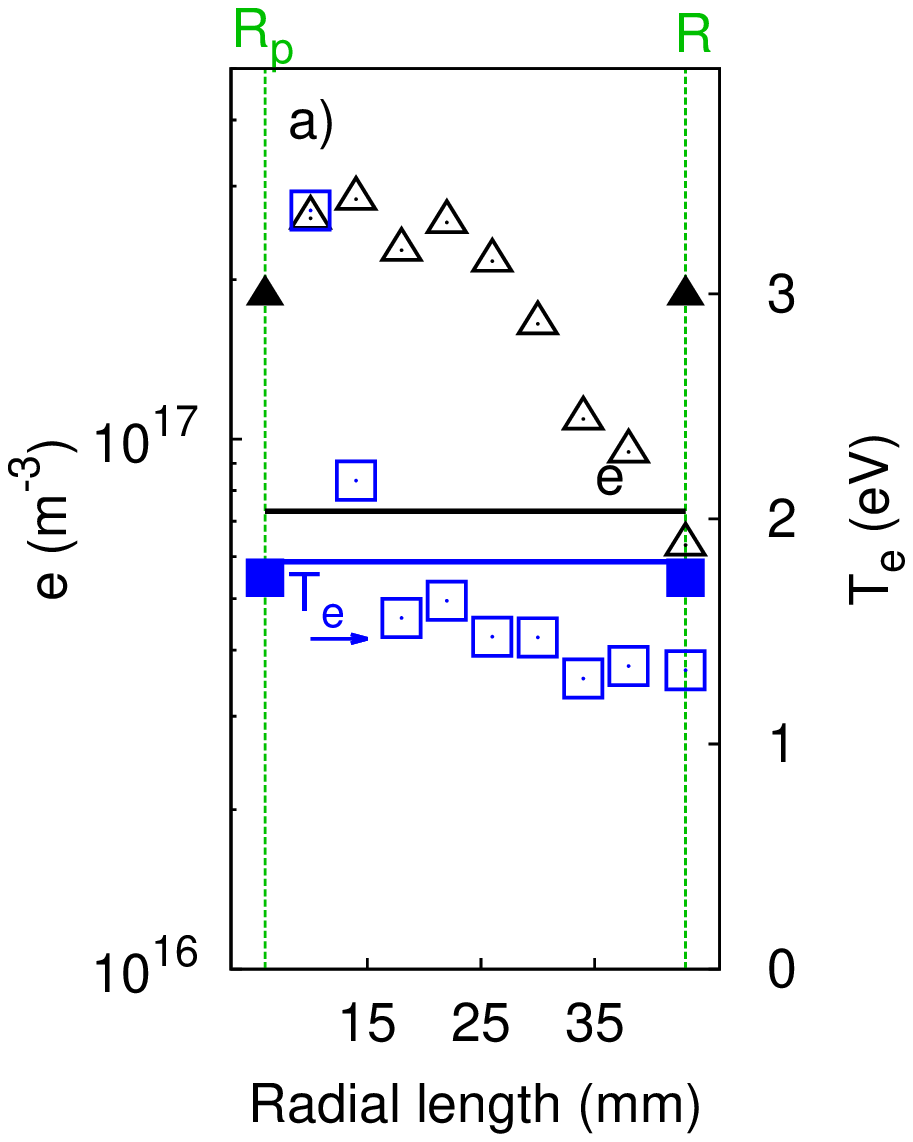}
\endminipage\hfill
\minipage{0.32\textwidth}
  \includegraphics[width=0.9\linewidth]{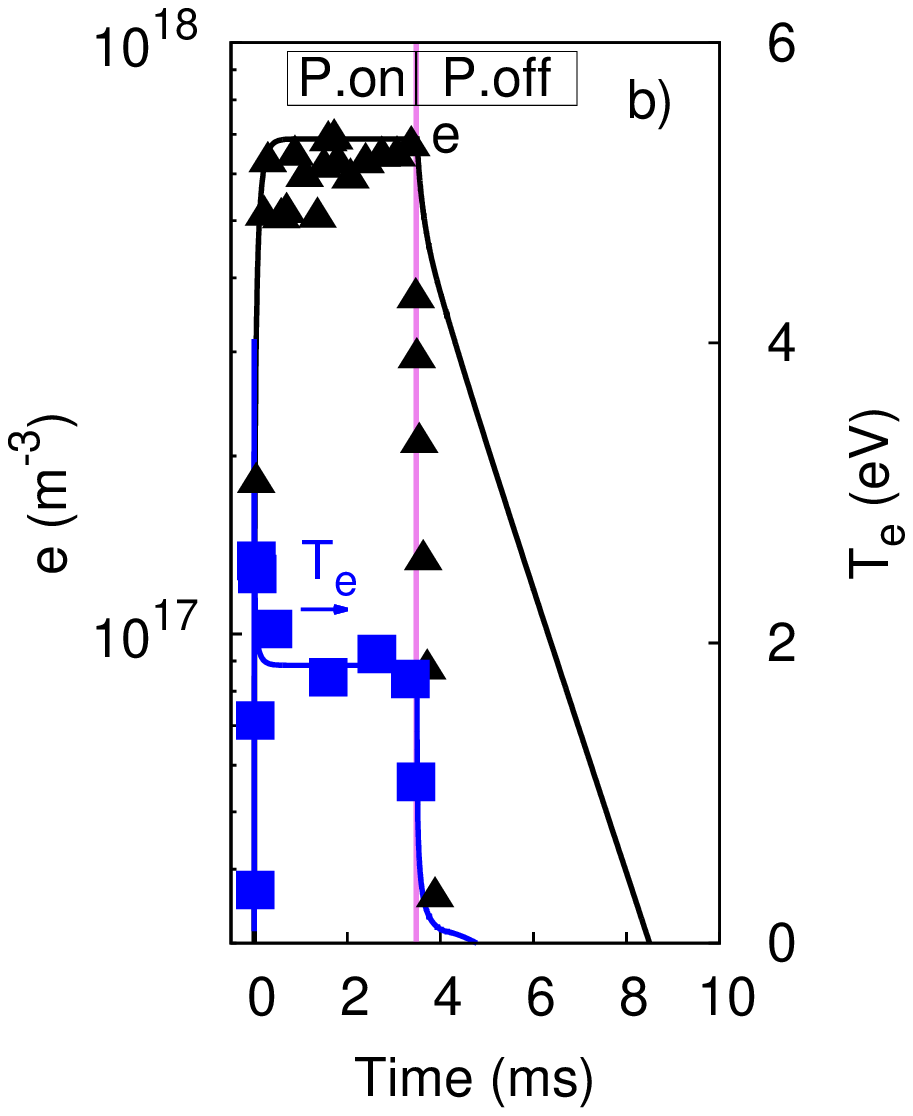}
\endminipage\hfill
\minipage{0.32\textwidth}
  \includegraphics[width=0.9\linewidth]{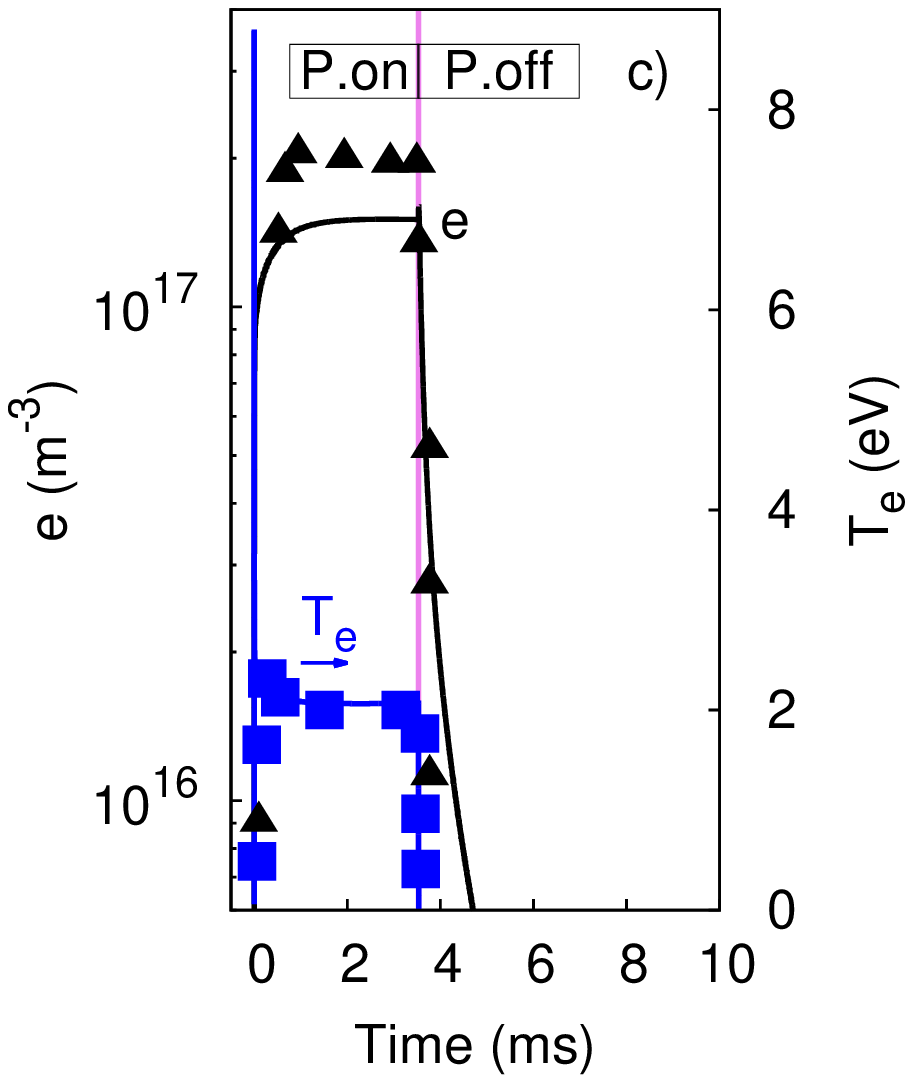}
\endminipage
\caption{The measurements of the discharge setup V - see table 
\ref{tab:sets} - and the simulation results (lines). 
The simulation results are volume-averaged and they are comparable to the 
spatially-averaged measurements.
a) Radially-resolved 
electron density ($\triangle$) and temperature (\textcolor{blue}{$\square$}) 
measurements as well as their radial averages (denoted by $\blacktriangle$ and 
\textcolor{blue}{$\blacksquare$}, respectively) on the stationary oxygen plasma. 
b) Time-resolved radially-averaged electron density ($\blacktriangle$) and 
temperature (\textcolor{blue}{$\blacksquare$}) measurements for a 
pulse-modulated power input in argon. c) Time-resolved radially-averaged 
measurements of electron density ($\blacktriangle$) and temperature 
(\textcolor{blue}{$\blacksquare$}) for a pulse-modulated power input in oxygen.}
\label{fig:comp_coapm}
\end{figure}
\twocolumngrid
\end{widetext}

The measurements on the discharge setup V of either argon or oxygen for 
continuous and pulse-modulated powers are shown in figure \ref{fig:comp_coapm} 
together with the simulation results. The radially-resolved and 
radially-averaged electron density and temperature in continuous 
power oxygen plasma are given in \ref{fig:comp_coapm}-a). The power input is $600\:$W 
at a pressure of $25\:$Pa and the gas temperature is $500\:$K with a gas flow 
rate of $200\:$sccm. The transient electron density and temperature 
of the pulse-modulated discharge in argon and oxygen are shown in figures 
\ref{fig:comp_coapm}-b) and \ref{fig:comp_coapm}-c), respectively. The pressure 
is $30\:$Pa for argon and $25\:$Pa for oxygen discharge both at a gas 
temperature of $900\:$K and the peak power input for both pulse-modulated 
measurements is $1500\:$W. The gas flow rates are $100\:$sccm in argon and 
$200\:$sccm in oxygen. The simulation results agree well with the 
radially-averaged stationary as well as transient measurements, however, the 
electron decay in the argon afterglow suggests considerably smaller decay rate 
compared to the experimental observation. The estimation of the afterglow 
edge-to-centre ratio to the unity - for example, after the electrons thermally 
equilibrate with the background gas - imposes much faster decay rate. However, 
such an assumption is not preferred due to the lack of data in the afterglow 
spatial structures, regarding validity requirements as well as the resultant 
sudden change in the decay rate. We also observe that the electron temperature 
decays much slowly as its value decreases mostly due to the insufficient loss 
mechanism.

Table \ref{tab:wal} shows the percentage of the wall loss in the net 
loss rate of certain species within the considered discharges.
The wall flux often forms a significant portion of the net loss in both 
cylindrical and coaxial structures. For the dominant ions, it reaches up to 
$98 \: \%$ for Ar$^+$ and up to $58 \: \%$ for O$_2^+$. 
Among the neutrals, the wall flux also considerably contributes to the net loss for O, 
$\text{O}_2(a^1 \Delta_g)$ and $\text{O}_2(b^1 \sum_g^+)$,
although, the maximum percentages are lower compared to those of the ions.
The wall loss rates of Ar(4s) and Ar(4p) are negligible, and hence,
the simulations are not sensitive to the wall de-excitation probabilities of 
the excited argon species.

Although the O wall recombination is an important loss mechanism, 
the recombination probability has only a weak effect on the simulations for a range 
of $0.009-1$. Because of this and due to lack of data, a recombination probability
identical to that of the quartz wall is assumed at the entire radial boundaries of the
coaxial structure V. Such a weak effect of the recombination probability is in contrast with the 
observations at low pressure oxygen plasma \cite{Lee95} that we also confirm 
with the same chemical kinetics of the current study \cite{Kem16}. The wall 
recombination is incomparably the most dominant mechanism both in the loss of 
oxygen atom and the production of the molecular oxygen at low pressure. 
However, in the relatively higher pressure regime considered, ozone-based reactions
also have a significant contribution to the oxygen atom loss and to the
molecular oxygen production. These reactions substantially suppress the 
critical role of the wall recombination probability on the surface wave discharges.

\onecolumngrid
\begin{widetext}
\begin{center}
\begin{table}[h!]\footnotesize
\begin{tabular}{lccc|lccc}
\\[\dimexpr-\normalbaselineskip+3pt]
\hline
\\[\dimexpr-\normalbaselineskip+3pt]
\\[\dimexpr-\normalbaselineskip+3pt]
Argon species   &           (I)            &       (IV)          &    (V)  &     Oxygen species    &    (II)  &  (V)  \\ 
\\[\dimexpr-\normalbaselineskip+3pt]                                                                                                                                                                            
\hline
\\[\dimexpr-\normalbaselineskip+3pt]                                                                                                                                                                            
$\text{Ar}^+$   &$  48  $ & $ 32 $  & $ 98 $ & $\text{O}_2^+$            &   $ 08 $ & $   58  $  \\ 
$\text{Ar(4sr)}$& $4 \times 10^{-3}$  & $3 \times 10^{-3}$  & $3 \times 10^{-2}$ & $\text{O}$                &   $ 10 $ & $   38  $  \\  
$\text{Ar(4p)}$ & $4 \times 10^{-4}$  & $4 \times 10^{-5}$  & $4 \times 10^{-4}$ & $\text{O}_2(a^1 \Delta_g)$&   $ 25 $ & $   29  $  \\ 
                &           &       &        & $\text{O}_2(b^1 \sum_g^+)$&   $ 07 $ & $   56  $  \\
\\[\dimexpr-\normalbaselineskip+3pt]                                                                                                                  
\hline
\end{tabular}
\caption{The percentage of the wall loss rate - either equations 
\ref{eqn:cyl_wal} for the cylindrical or equations \ref{eqn:coa_wal} for the 
coaxial structures - in the net loss rate of dominant ions and certain neutrals.
Here, we only consider the discharges operated with continuous power input and 
refer to table \ref{tab:sets} for the corresponding operation parameters. 
We show the calculation results of setup II at a power input of $107\:$W and at a 
pressure of $133\:$Pa.
} 
\label{tab:wal}
\end{table}
\end{center}
\end{widetext}
\twocolumngrid

We observe that the considered oxygen discharges are attachment dominated with 
$\tau_{an} K_a \geq 20$, hence the negative ions are not in Boltzmann 
equilibrium. On the other hand, the simulation results are not sensitive to the 
resultant variation in the Bohm velocity. The negative ion loss mechanism is 
dominated by the detachment as expected in the considered pressure regime 
\cite{Fra02}. A hypothetical inclusion of an axial ion edge-to-centre ratio, 
for example, with the critical edge electron density of the surface-wave 
propagation \cite{Jau12}, are negligible on the results.

The analytically estimated radial Bessel profile is also proposed earlier in 
cylindrical argon \cite{Jim12} and oxygen discharges of opposite asymptotic $P$ 
limits \cite{Fer88,Dan90,Dan89}. The former should be disturbed due to 
molecular-assisted recombination in the considered case of operation conditions.
A Bessel radial distribution in coaxial argon discharge is also observed 
\cite{Rah14_1}, however, the profile transforms to a similar Gaussian-like shape 
as the molecular-assisted recombination dominates over the diffusion loss. 
The simulation results suggest that the molecular-assisted recombination 
dominates in setup IV, whereas the diffusive wall loss dominates in setup V. The 
measurements of coaxial argon and oxygen discharges suggest a profile peak 
closer to the inner wall compared to the analytical estimation since the power 
density is localized near the inner wall.

The ion edge-to-centre ratio is numerically comparable with the 
low-collisionality analogue
($h^+_{iR}=3.90 \times 10^{-3}$ and $\bar{h}^+_{iR}=3.88 \times 10^{-3}$)
in argon surfatron (I) and the positive ions are volume loss dominated with a 
comparable wall loss rate. The value of the neutral edge-to-centre ratio is also 
similar with the earlier derivation for the excited argon species 
($h^N_{iR}=1.73 \times 10^{-3}$ and $\bar{h}^N_{iR}=2.50 \times 10^{-3}$). The 
difference in the numerical values is due to the factor of $(2/2.405)^2$ and 
the addition of the axial walls are negligible on the numerical difference. The 
interchange of the ion and the neutral edge-to-centre ratios with the earlier 
analogues are negligible on the simulation results.

The numerical values of the ion edge-to-centre ratios significantly differ in 
the case of oxygen surfatron (II) due to the single electronegative region 
flat-topped term $\bar{h}^+_{iR,c}$. The ratio $\bar{h}^+_{iR}$ is higher than 
$h^+_{iR}$ mostly by a factor of about $10$ for a volume loss dominated 
positive ions within a range of degree of the electronegativity $2-5$. The 
interchange of the ratio with $\bar{h}^+_{iR}$ in the simulations reduces a 
variety of particle densities and raises the electron temperature. This also 
changes the dominant positive ion loss mechanism to the diffusive wall losses 
and increases the degree of electronegativity to the range $4-7$. The numerical 
density values are still in the same order and the usage of any edge-to-centre 
ratios do not lead better agreement with the considered measurements. 
Compared with the earlier assumption of about $0.05$ in the edge-to-volume-averaged 
formalism \cite{Kem15}, the derived edge-to-centre ratio is smaller by a factor of 
about $0.17$. 
A factor of difference between the edge-to-centre and edge-to-volume-averaged formalisms
is about $0.5$.
The neutral 
edge-to-centre ratios $h^N_{iR}$ and $\bar{h}^N_{iR}$ are similar in their 
numerical values and their interchange with each other is not effective on the 
simulation results. The difference between the ion edge-to-centre ratios is 
even larger in setup III. The ratio $\bar{h}^+_{iR}$ is larger than $h^+_{iR}$ 
with a factor of about $200$ for $\alpha \approx 12$. 
However, the positive ions are volume loss dominated and the 
interchange of the edge-to-centre ratios do not significantly alter the 
simulation results.

%% file: conc.tex
\section{Conclusions} 
\label{sec:con}
 
A volume-averaged global model is developed for the surface-wave discharges of 
cylindrical and coaxial structures for both the electropositive and the 
electronegative plasmas. A set of edge-to-centre density ratios for the neutrals 
and the positive ions are derived based on an analytical one-dimensional model 
in order to estimate regarding wall loss rates. The analytical model of charged
species is based on the three-component plasma with an assumption of constant or 
piecewise constant degree of electronegativity as suggested by a variety of earlier 
studies \cite{Fer83,Fra99,Kim06} in the considered collisionality regime. The 
simulation results are compared with various volume-averaged measurements, and 
a good agreement is obtained in argon or oxygen surface-wave discharges of 
cylindrical and coaxial structures for both continuous and pulse-modulated 
power input. The calculations show that the wall flux is often an important 
loss mechanism. A numerical comparison is made between the derived edge-to-centre 
ratios in the cylindrical structure and their analogues that are used in the 
low-pressure plasmas. A difference is observed only in the electronegative 
oxygen plasma, where the validity of the low-pressure analogue is expired 
\cite{Fra99,Fra01,Fra02}.

The inhomogeneity in the discharge is often significantly localized, for 
example, inside the surfatron launcher, near the dielectric surface. However,
a homogeneity assumption with a zeroth order degree of homogeneity still holds 
in the volume-averaged quantities within the numerical range of the model 
sensitivity. Analytically estimated profiles are in approximate agreement with 
the earlier studies with a maximum located more radially outward in the coaxial 
structure compared to measurements. The resultant edge-to-centre ratios 
sufficiently describe the wall flux rates for an experimentally confirmed 
volume-averaged particle densities within the sensitivity of the model. 
The power transfer efficiency of the coaxial discharge of interest (V) is only 
estimated ignoring the small portion of the axial length with a smaller outer 
radius. A more accurate estimation can be acquired by a coupling of the 
microwave propagation with an appropriate spatially-resolved plasma model. The 
chemical sets are adopted from earlier studies in literature and a throughout 
analysis of the reaction mechanisms is still required for further accuracy. 

\section*{Acknowledgements}
The authors gratefully acknowledge the support by the Federal Ministry of 
Education and Research (BMBF) within the framework of the project PluTO+
and by DFG (German Research Foundation) within the framework of the 
Sonderforschungsbereich SFB-TR 87. The authors also thank Simon H{\"u}bner
for sharing his measurements and Thomas Gudmundsson for his review and precious 
comments on the final draft. 

%% file: appendix.tex
\newpage
\clearpage
{\bf \LARGE Appendix}
\appendix
\section{Derivation of ion edge-to-centre ratios}
\label{sec:app}

The general solution of the ions is earlier given in the form 
\begin{equation}
n^+_i(r) = C_1 J_0 \left( \chi r/R \right) + C_2 Y_0 \left( \chi r/R\right),
\label{eqn:app_sol}
\end{equation}
where the boundary conditions and 
the normalization of the maximum to 
the unity define the special - dimensionless - form. 
At any Bohm point $r'_B$ that is in the vicinity of the 
corresponding radial plasma wall $R'$, the ions satisfy
\begin{equation} 
-D_{i+} \frac{\partial n^+_i}{\partial r} \left. \right|_{r'_B} = (\pm) u_{+,BE} n^+_i \left. \right|_{r'_B}, 
\label{eqn:app_bnd}
\end{equation} 
where ${r'_B}$ corresponds to either the single radial Bohm point of the cylindrical 
($R'=R$) or double radial Bohm points of the coaxial structure 
($R'=R$ and $R'=R_p$) and only the inner wall ($R'=R_p$) incorporates the 
negative sign (-) on the right-hand side. The edge-to-centre ratio at any radial 
wall is 
\cite{LieBook2005}
\begin{equation} 
h^+_{iR'} = \frac{1}{1+\alpha} (u_{BE}/u_B)n^+_i \left. \right|_{r'_B},
\end{equation}
where the centre value - corresponds to the maximum value for the coaxial structure - 
is unity via normalisation, $\frac{1}{1+\alpha}$ is an additional factor of 
normalization in the electronegative discharges \cite{Lee95} and $(u_{BE}/u_B)$
is the effect of the electronegativity on the Bohm velocity. Assuming negligibly 
small sheath length , $\left| r'_B - R'\right|/R \gg 1$, the Taylor series 
expansion at $R'$ leads 
\begin{equation}
\frac{\partial n^+_i}{\partial r} \left. \right|_{r'_B} \approx \frac{\partial n^+_i}{\partial r} \left. \right|_{R'}.
\end{equation}
Hence, using this with equation \ref{eqn:app_bnd}, 
$n^+_i \left. \right|_{r'_B}$ can be approximated by the following relation
\begin{equation}                                                                
n^+_i \left. \right|_{r'_B} \approx (\mp)                                          
\frac{ \chi D_+}{R u_{+,BE}} (C_1 J_1 \left( \chi R'/R \right) + C_2 Y_1 \left( \chi R'/R\right)).
\label{eqn:app_1}
\end{equation} 
However, it is observed that the edge-to-centre ratios via this approximation
often cause computationally unstable simulations. A numerically equivalent and 
computationally more stable alternative can be estimated from this approximation 
in the following form
\begin{widetext}
\onecolumngrid
\begin{equation}                                                                
n^+_i \left. \right|_{r'_B} \approx 
\left( 1 + \left( \frac{R u_{+,BE}}{\chi D_+ \left(C_1 J_1 \left( \chi R'/R \right) + C_2 Y_1 \left( \chi R'/R\right)\right)}\right)^2 \right)^{-1/2}.
\label{eqn:app_2}
\end{equation}
\end{widetext}
\twocolumngrid
For the sake of computational stability, we prefer the latter form with the 
edge-to-centre ratio
\begin{widetext}
\onecolumngrid
\begin{equation}
h^+_{iR'} \approx  \frac{1}{1+\alpha} (u_{BE}/u_B) \left( 1 + \left( \frac{R u_{+,BE}}{\chi D_{i+} 
\left( C_1 J_1 \left( \chi R'/R \right) + C_2 Y_1 \left( \chi R'/R\right) \right)
}\right)^2 \right)^{-1/2}.
\end{equation}
\end{widetext}
\twocolumngrid

%% file: chemkin.tex
\newpage

\section{Chemical kinetics}
\label{sec:app2}

Chemical sets by Gudmundsson {\em et al} \cite{Gud07} are adopted in argon 
plasma simulations as well as argon excimer reactions and three body 
recombination by Jimenez {\em et al} \cite{Jim12} and Kabouzi {\em et al} 
\cite{Kab07}. A previously adopted species and reaction sets are used in the 
oxygen simulations without a reduction \cite{Kem15}.

\onecolumngrid
\begin{center}
\begin{figure}[htbp!]
\resizebox{0.7\textwidth}{0.7\textwidth}{
\begin{tikzpicture} 
\large

\draw[>=latex,->, ultra thick] (-1.2,-1) -- (-1.2,17) node[above] {E (eV)};
\draw[>=latex,->, dotted] (-1.5,0) -- (17,0) node[left] { };

\draw[level] (0.3,1.1) -- node[left=15pt] {1.1} node[right=15pt] {O$^-$} (1,1.1);
\draw[level] (0.3,2.56) -- node[left=15pt] {2.56} node[right=15pt] {O($^3 P$)} (1,2.56);
\draw[level] (0.3,4.52) -- node[left=15pt] {4.52} node[right=15pt] {O($^1 D$)} (1,4.52);
\draw[level] (0.3,16.16) -- node[left=15pt] {16.16} node[right=15pt] {O$^+$} (1,16.16);

\draw[level] (4.5,0) -- node[left=15pt] {0} node[right=15pt] {O$_2$} (5.5,0);

\draw[level] (4.5,-0.45) -- node[left=15pt] {-0.45} node[right=15pt] {O$_2^-$} (5.5,-0.45);
\draw[level] (6,0.98) -- node[left=15pt] {0.98} node[right=15pt] {O$_2 (a^1 \Delta_g)$} (7,0.98);
\draw[level] (6,1.63) -- node[left=15pt] {1.63} node[right=15pt] {O$_2 (b^1 \sum_g^+)$} (7,1.63);
\draw[level] (6,4.2) -- node[left=15pt] {4.2} node[right=15pt] {O$_2 (A^3 \sum_u^+, A^3 \Delta_u, c^1 \sum_u^{-})$} (7,4.2);
\draw[level] (4.5,12.1) -- node[left=15pt] {12.1} node[right=15pt] {O$_2^+$} (5.5,12.1);

\draw[level] (11,-0.64) -- node[left=15pt] {-0.64} node[right=15pt] {O$_3^-$} (12,-0.64);
\draw[level] (11,1.46) -- node[left=15pt] {1.46} node[right=15pt] {O$_3$} (12,1.46);

\draw[level] (13,0) -- node[left=15pt] {0} node[right=15pt] {Ar} (14,0);

\draw[level] (13,11.65) -- node[left=15pt] {11.65} node[right=15pt] {Ar4s[r,m]} (14,11.65);

\draw[level] (13,13.17) -- node[left=15pt] {13.17} node[right=15pt] {Ar4p} (14,13.17);

\draw[level] (13,15.76) -- node[left=15pt] {15.76} node[right=15pt] {Ar$^+$} (14,15.76);

\draw[level] (12,15.01) -- node[left=15pt] {15.01} node[right=15pt] {Ar$_2^+$} (13,15.01);

\end{tikzpicture}}
 \caption{Energy level diagram of Ar and $\text{O}_2$ plasma species.}
\label{fig:o2energydiagram}
\end{figure}
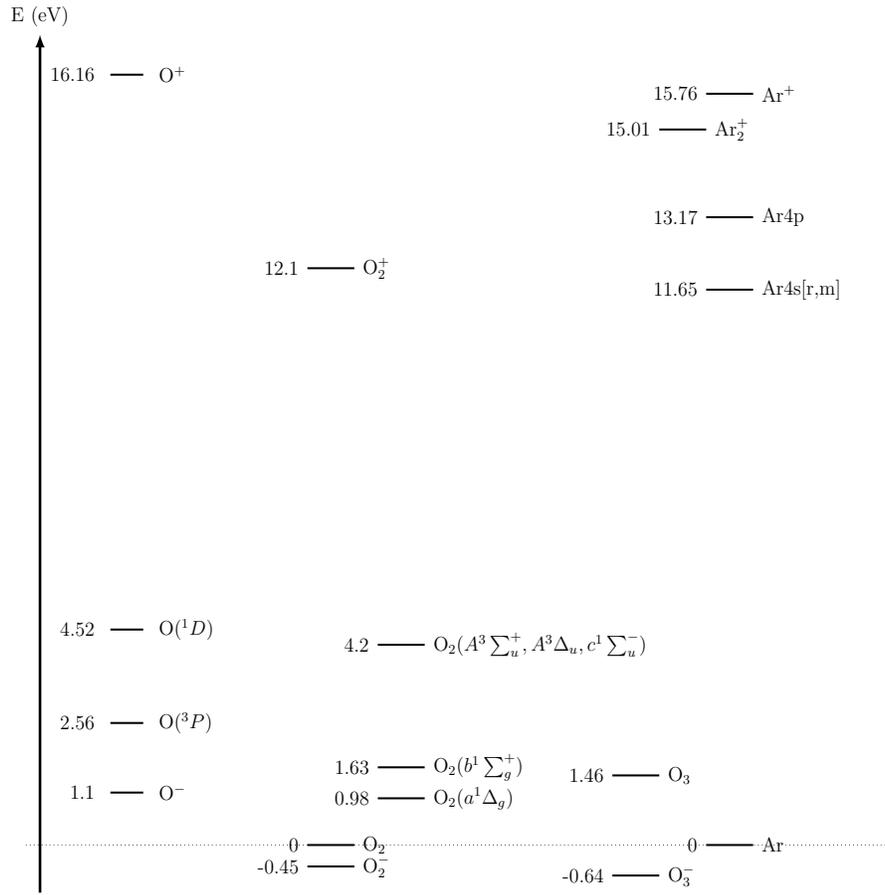
\end{center}

\begin{table}\scriptsize
\centering
\caption{Argon plasma reactions. We adopt the chemical set by Gudmundsson 
{\em et al} \cite{Gud07} and  additionally include the argon excimer ion kinetics 
as well as three-body recombination by Jimenez {\em et al} \cite{Jim12} or 
Kabouzi {\em et al} \cite{Kab07}. The units of the rate coefficients are 
$\text{m}^3\text{s}^{-1}$ and those of electron and gas temperatures are eV and 
K, respectively, if not stated otherwise.}
\begin{tabular}{p{0.7cm}p{4.5cm}p{8.4cm}p{1.0cm}p{0.3cm}}
\\[\dimexpr-\normalbaselineskip+3pt]
\hline
\\[\dimexpr-\normalbaselineskip+3pt]
\# &Reaction     & Rate Coefficient & Ref & \\ \hline
\\[\dimexpr-\normalbaselineskip+3pt]
1  & $ e + \text{Ar} \rightarrow \text{Ar}^+ + 2e                       $ & $ 2.3 \times 10^{-14} T_e^{0.59} e^{-17.44/T_e}$  &  \cite{Str95} &     \\ 
\\[\dimexpr-\normalbaselineskip+3pt]
2  & $ e + \text{Ar} \rightarrow e + \text{Ar}(4s_m)    $ & $ 5.0 \times 10^{-15}  e^{-12.64/T_e}$  &  \cite{Tac86} &     \\ 
\\[\dimexpr-\normalbaselineskip+3pt]
3  & $ e + \text{Ar} \rightarrow e + \text{Ar}(4s_m)    $ & $ 1.4 \times 10^{-15}  e^{-12.42/T_e}$  &  \cite{Tac86} &     \\ 
\\[\dimexpr-\normalbaselineskip+3pt]
4  & $ e + \text{Ar} \rightarrow e + \text{Ar}(4s_r)    $ & $ 1.9 \times 10^{-15}  e^{-12.60/T_e}$  &  \cite{Tac86} &     \\ 
\\[\dimexpr-\normalbaselineskip+3pt]
5  & $ e + \text{Ar} \rightarrow e + \text{Ar}(4s_r)    $ & $ 2.7 \times 10^{-16}  e^{-12.14/T_e}$  &  \cite{Tac86} &     \\ 
\\[\dimexpr-\normalbaselineskip+3pt]
6  & $ e + \text{Ar} \rightarrow e + \text{Ar}(4p)      $ & $ 2.1 \times 10^{-14}  e^{-13.13/T_e}$  &  \cite{Egg75} &     \\ 
\\[\dimexpr-\normalbaselineskip+3pt]
7  & $ e + \text{Ar}(4s_m) \rightarrow e + \text{Ar}    $ & $ 4.3 \times 10^{-16}  T_e^{0.74}    $  &  \cite{Ash95} &     \\ 
\\[\dimexpr-\normalbaselineskip+3pt]
8  & $ e + \text{Ar}(4s_m) \rightarrow \text{Ar}^+ + 2e $ & $ 6.8 \times 10^{-15}  T_e^{0.67} e^{-4.20/T_e}$  &  \cite{Kan85} &     \\ 
\\[\dimexpr-\normalbaselineskip+3pt]
9  & $ e + \text{Ar}(4s_m) \rightarrow e + \text{Ar}(4s_r) $ & $ 3.7 \times 10^{-13}             $  &  \cite{Fer85} &     \\ 
\\[\dimexpr-\normalbaselineskip+3pt]
10  & $ e + \text{Ar}(4s_m) \rightarrow e + \text{Ar}(4p)  $ & $ 8.9 \times 10^{-13} T_e^{0.51} e^{-1.59/T_e}$  &  \cite{Kan85} &     \\ 
\\[\dimexpr-\normalbaselineskip+3pt]
11  & $ 2 \text{Ar}(4s_m) \rightarrow 2 \text{Ar}    $ & $ 2.0 \times 10^{-13}                   $  &  &     \\ 
\\[\dimexpr-\normalbaselineskip+3pt]
12?  & $ \text{Ar} + \text{Ar}^+ \rightarrow \text{Ar} + \text{Ar}^+    $ & $ 2.2 \times 10^{-16}$  &  \cite{Ney67} &     \\ 
\\[\dimexpr-\normalbaselineskip+3pt]
13  & $ \text{Ar}(4s_m) + \text{Ar}(4s_r) \rightarrow \text{Ar} + \text{Ar}^+ + e     $ & $ 2.1 \times 10^{-15} $  &  \cite{Bas94} &     \\ 
\\[\dimexpr-\normalbaselineskip+3pt]
14  & $ \text{Ar}(4p) + \text{Ar}(4p) \rightarrow \text{Ar} + \text{Ar}^+ + e     $ & $ 5.0 \times 10^{-16}     $  &  \cite{Kan85} &     \\ 
\\[\dimexpr-\normalbaselineskip+3pt]
14  & $ \text{Ar}(4s_m) + \text{Ar}(4s_m) \rightarrow \text{Ar} + \text{Ar}^+ + e $ & $ 6.4 \times 10^{-16}     $  &  \cite{Fer85} &     \\ 
\\[\dimexpr-\normalbaselineskip+3pt]
15  & $ e + \text{Ar}(4p) \rightarrow \text{Ar}^+ + 2e     $ & $ 1.8 \times 10^{-13} T_e^{0.61} e^{-2.61/T_e}   $  &  \cite{Kan85} &     \\ 
\\[\dimexpr-\normalbaselineskip+3pt]
16  & $ e + \text{Ar}(4p) \rightarrow \text{Ar}(4s_r) + e  $ & $ 3.0 \times 10^{-13} T_e^{0.51}                 $  &  \cite{Ash95} &     \\ 
\\[\dimexpr-\normalbaselineskip+3pt]
17  & $ e + \text{Ar}(4p) \rightarrow \text{Ar}(4s_m) + e  $ & $ 3.0 \times 10^{-13} T_e^{0.51}                 $  &  \cite{Ash95} &     \\ 
\\[\dimexpr-\normalbaselineskip+3pt]
18  & $ e + \text{Ar}(4p) \rightarrow \text{Ar} + e        $ & $ 3.9 \times 10^{-16} T_e^{0.71}                 $  &  \cite{Ash95} &     \\ 
\\[\dimexpr-\normalbaselineskip+3pt]
19  & $ \text{Ar} + \text{Ar}(4s_m) \rightarrow 2 \text{Ar}$ & $ 2.1 \times 10^{-21}                            $  &  \cite{Bas94} &     \\ 
\\[\dimexpr-\normalbaselineskip+3pt]
20  & $ e + \text{Ar}(4s_r) \rightarrow \text{Ar} + e      $ & $ 4.3 \times 10^{-16} T_e^{0.74}                 $  &  \cite{Ash95} &     \\ 
\\[\dimexpr-\normalbaselineskip+3pt]
21  & $ e + \text{Ar}(4s_r) \rightarrow \text{Ar}(4s_m) + e$ & $ 9.1 \times 10^{-13}                            $  &  \cite{Fer85} &     \\ 
\\[\dimexpr-\normalbaselineskip+3pt]
22  & $ e + \text{Ar}(4s_r) \rightarrow \text{Ar}(4p) + e  $ & $ 8.9 \times 10^{-13} T_e^{0.51} e^{-1.59/T_e}   $  &  \cite{Kan85} &     \\ 
\\[\dimexpr-\normalbaselineskip+3pt]
23  & $ \text{Ar}(4s_r) \rightarrow \text{Ar}              $ & $ 1.0 \times 10^{05} \:                        1/$s &  \cite{Hur74} &     \\ 
\\[\dimexpr-\normalbaselineskip+3pt]
24  & $ \text{Ar}(4p) \rightarrow \text{Ar}                $ & $ 3.2 \times 10^{07} \:                        1/$s &  \cite{Ash95} &     \\ 
\\[\dimexpr-\normalbaselineskip+3pt]
25  & $ \text{Ar}(4p) \rightarrow \text{Ar}(4s_m)          $ & $ 3.0 \times 10^{07} \:                        1/$s &  \cite{Lee06} &     \\ 
\\[\dimexpr-\normalbaselineskip+3pt]
26  & $ \text{Ar}(4p) \rightarrow \text{Ar}(4s_r)          $ & $ 3.0 \times 10^{07} \:                        1/$s &  \cite{Lee06} &     \\ 
\\[\dimexpr-\normalbaselineskip+3pt]
27  & $ \text{Ar}_2^+ + e \rightarrow \text{Ar}(4s_m) + \text{Ar}      $ & $ 1.04 \times 10^{-12} (T_e[K]/300)^{-0.67} (1 - e^{-418/T_h})/(1-0.31 e^{-418/T_h})$  &  \cite{Kab07} &     \\ 
\\[\dimexpr-\normalbaselineskip+3pt]
28  & $ \text{Ar}_2^+ + e \rightarrow \text{Ar}(4s_r) + \text{Ar}      $ & $ 1.04 \times 10^{-12} (T_e[K]/300)^{-0.67} (1-e^{-418/T_h})/(1-0.31 e^{-418/T_h})$  &  \cite{Kab07} &     \\ 
\\[\dimexpr-\normalbaselineskip+3pt]
29  & $ \text{Ar}_2^+ + e \rightarrow \text{Ar}^+ + \text{Ar} + e      $ & $ 1.11 \times 10^{-12} e^{-(2.94-3 (T_h [eV] - 0.026))/T_e}$  &  \cite{Kab07} &     \\ 
\\[\dimexpr-\normalbaselineskip+3pt]
30  & $ \text{Ar}^+ + 2 \text{Ar} \rightarrow \text{Ar}_2^+ + \text{Ar}$ & $ 2.25 \times 10^{-43} (T_h/300)^{-0.4} $  &  \cite{Kab07} &     \\ 
\\[\dimexpr-\normalbaselineskip+3pt]
31  & $ \text{Ar}_2^+ + \text{Ar} \rightarrow \text{Ar}^+ + 2\text{Ar} $ & $ 5.22 \times 10^{-16} (T_h[eV])^{-1.0} e^{-1.304/T_h[eV]}$  &  \cite{Kab07} &     \\ 
\\[\dimexpr-\normalbaselineskip+3pt]
32  & $ \text{Ar}^+ + 2 e \rightarrow \text{Ar}(4p) + e                $ & $ 5.00 \times 10^{-39} (T_e)^{-4.5}               $  &  \cite{Bho04} &     \\ 
\\[\dimexpr-\normalbaselineskip+3pt]
33  & $ \text{Ar}^+ + 2 e \rightarrow \text{Ar} + e                    $ & $ 8.75 \times 10^{-39} T_e^{-2.25}                 $  &  \cite{Rai91} &     \\ 
\\[\dimexpr-\normalbaselineskip+3pt]
\hline 
\end{tabular}
\label{tab:rear}
\end{table}

\begin{table}\scriptsize
\centering
\caption{Electron - Oxygen reactions \cite{Kem15}. The rate coefficients are 
given in $\text{m}^3\text{s}^{-1}$ and the electron temperature is given in eV.
The symbol ``*'' is used to specify the inverse reaction coefficients calculated 
by {\em detailed balancing}.}
\begin{tabular}{p{0.7cm}p{6.0cm}p{5.5cm}p{2.0cm}p{0.9cm}}
\\[\dimexpr-\normalbaselineskip+3pt]
\hline
\\[\dimexpr-\normalbaselineskip+3pt]
\# &Reaction     & Rate Coefficient & Ref & \\ \hline
\\[\dimexpr-\normalbaselineskip+3pt]
1  & $ e + \text{O}_2 \rightarrow \text{O}_2^+ + 2e                       $ & $ 2.01 \times 10^{-15} T_e^{1.09} e^{-12.41/T_e}$  &  \cite{Str96} &     \\ 
\\[\dimexpr-\normalbaselineskip+3pt]
2  & $ e + \text{O}_2 \rightarrow \text{O}(^3P) + \text{O}^+ + 2e                $ & $ 1.04 \times 10^{-14} T_e^{1.11} e^{-21.27/T_e}$  &  \cite{Str96} &     \\ 
\\[\dimexpr-\normalbaselineskip+3pt]
3  & $ e + \text{O}_2 \rightarrow \text{O}(^3P) + \text{O}^{ - }                 $ & $ 1.12 \times 10^{-15} T_e^{-1.41} e^{-6.16/T_e} $  &  \cite{Jaf92} &   \\ 
\\[\dimexpr-\normalbaselineskip+3pt]
4  & $ e + \text{O}_2 (a^1 \Delta_g) \rightarrow \text{O}(^3P) + \text{O}^{ - }  $ & $ 4.33 \times 10^{-15} T_e^{-1.39} e^{-5.17/T_e} $  &  \cite{Jaf92} &     \\ 
\\[\dimexpr-\normalbaselineskip+3pt]
5  & $ e + \text{O}_2 (a^1 \Delta_g) \rightarrow \text{O}(^1D) + \text{O}^{ - }  $ & $ 1.01 \times 10^{-15} T_e^{-1.46} e^{-7.36/T_e} $  &  \cite{Jaf92} &     \\ 
\\[\dimexpr-\normalbaselineskip+3pt]
6  & $ e + \text{O}_2 (A^3 \sum_u^+,...) \rightarrow \text{O}(^3P)+\text{O}^{ - } $ & $ 5.77 \times 10^{-16} T_e^{-0.90} e^{-0.12/T_e}$  &  \cite{Hay99} &   \\ 
\\[\dimexpr-\normalbaselineskip+3pt]
7  & $ e + \text{O}(^3P) \rightarrow \text{O}^+ + 2e                        $ & $ 4.75 \times 10^{-15} T_e^{0.78} e^{-14.27/T_e}$  &  \cite{Kim02} &     \\ 
\\[\dimexpr-\normalbaselineskip+3pt]
8  & $ e + \text{O}^{ - } \rightarrow \text{O}(^3P) + 2e                    $ & $ 4.64 \times 10^{-14} T_e^{0.50} e^{-3.44/T_e} $  &  \cite{Deu03} &   \\ 
\\[\dimexpr-\normalbaselineskip+3pt]
9  & $ e + \text{O}_2 \rightarrow \text{O}(^3P) + \text{O}(^1D) + e                $ & $ 8.45 \times 10^{-15} T_e^{0.38} e^{-11.84/T_e}$  &  \cite{Cos93} &     \\ 
\\[\dimexpr-\normalbaselineskip+3pt]
10 & $ e + \text{O}_2 \rightarrow \text{O}(^3P) + \text{O}(^3P) + e                $ & $ 9.49 \times 10^{-16} T_e^{0.38} e^{-11.84/T_e}$  &  \cite{Cos93} &     \\ 
\\[\dimexpr-\normalbaselineskip+3pt]
11 & $ e + \text{O}_2 \rightarrow \text{O}(^1D) + \text{O}(^1D) + e                $ & $ 9.49 \times 10^{-17} T_e^{0.38} e^{-11.84/T_e}$  &  \cite{Cos93} &     \\ 
\\[\dimexpr-\normalbaselineskip+3pt]
12 & $ e + \text{O}_2 \rightarrow \text{O}^+ + \text{O}^{ - } + e                  $ & $ 4.12 \times 10^{-14} T_e^{-0.25} e^{-20.80/T_e}$  &  \cite{Rap65} &     \\ 
\\[\dimexpr-\normalbaselineskip+3pt]
13 & $ e + \text{O}(^3P) \rightarrow \text{O}(^1D) + e                      $ & $ 2.19 \times 10^{-14} T_e^{-0.57} e^{-4.10/T_e} $  &  \cite{Doe92} &  * \\ 
\\[\dimexpr-\normalbaselineskip+3pt]
14 & $ e + \text{O}_2 \rightarrow \text{O}_2 (a^1 \Delta_g) + e             $ & $ 1.25 \times 10^{-14} T_e^{-0.97} e^{-5.51/T_e} $  &  \cite{Shy93} &  * \\ 
\\[\dimexpr-\normalbaselineskip+3pt]
15 & $ e + \text{O}_2 \rightarrow \text{O}_2 (b^1 \sum_g^+) + e             $ & $ 3.84 \times 10^{-15} T_e^{-1.05} e^{-5.71/T_e} $  &  \cite{Shy93} &  * \\ 
\\[\dimexpr-\normalbaselineskip+3pt]
16 & $ e + \text{O}_2 \rightarrow \text{O}_2 (A^3 \sum_u^+,...)+e $ & $ 2.39 \times 10^{-14} T_e^{-1.00} e^{-9.52/T_e}    $  &  \cite{Shy00} &  * \\ 
\\[\dimexpr-\normalbaselineskip+3pt]
17 & $ e + \text{O}_2 (a^1 \Delta_g) \rightarrow \text{O}_2 (b^1 \sum_g^+) + e   $ & $ 6.69 \times 10^{-15} T_e^{-0.56} e^{-1.73/T_e}    $  &  \cite{Tas06} &  * \\ 
\\[\dimexpr-\normalbaselineskip+3pt]
18 & $ e + \text{O}_2 (a^1 \Delta_g) \rightarrow \text{O}_2(A^3 \sum_u^+ ,...)+e$ & $7.23\times 10^{-14}T_e^{-1.25} e^{-7.63/T_e}$  &  \cite{Tas06} &  * \\ 
\\[\dimexpr-\normalbaselineskip+3pt]
19 & $ e + \text{O}_2 (b^1 \sum_g^+) \rightarrow \text{O}_2(A^3 \sum_u^+ ,...)+e$ & $8.47\times 10^{-14} T_e^{-1.23} e^{-7.12/T_e} $  &  \cite{Tas06} &  * \\
\\[\dimexpr-\normalbaselineskip+3pt]
20 & $ e + \text{O}_3 \rightarrow \text{O}_2 + \text{O}^{ - }                     $ & $ 3.24 \times 10^{-15} T_e^{-0.94} e^{-0.91/T_e} $  &  \cite{Ran99} & \\ 
\\[\dimexpr-\normalbaselineskip+3pt]
21 & $ e + \text{O}_3 \rightarrow \text{O}(^3P) + \text{O}_2^{ - }                $ & $ 9.56 \times 10^{-16} T_e^{-1.26} e^{-0.95/T_e} $  &  \cite{Ran99} & \\ 
\\[\dimexpr-\normalbaselineskip+3pt]
22 & $ e + \text{O}_3 \rightarrow \text{O}_2 + \text{O}(^3P) + e                  $ & $ 1.42 \times 10^{-14} T_e^{-0.68} e^{-2.60/T_e} $  &  \cite{Gup05} & \\ 
\\[\dimexpr-\normalbaselineskip+3pt]
23 & $ e + \text{O}_3 \rightarrow \text{O}_2 (a^1 \Delta_g) + \text{O}(^3P) + e   $ & $ 4.16 \times 10^{-15} T_e^{-0.73} e^{-3.15/T_e} $  &  \cite{Gup05} & \\ 
\\[\dimexpr-\normalbaselineskip+3pt]
24 & $ e + \text{O}_3 \rightarrow \text{O}_2 (a^1 \Delta_g) + \text{O}(^1D) + e   $ & $ 6.68 \times 10^{-15} T_e^{-0.82} e^{-4.07/T_e} $  &  \cite{Gup05} & \\ 
\\[\dimexpr-\normalbaselineskip+3pt]
25 & $ e + \text{O}_3 \rightarrow \text{O}_2 (b^1 \sum_g^+) + \text{O}(^1D) + e   $ & $ 1.34 \times 10^{-13}T_e^{-0.87} e^{-6.63/T_e} $  &  \cite{Gup05} & \\ 
\\[\dimexpr-\normalbaselineskip+3pt]
26 & $ e + \text{O}_2^{ - } \rightarrow \text{O}_2 + 2e                    $ & $ 5.74 \times 10^{-14} T_e^{0.58} e^{-4.68/T_e} $  &  \cite{Deu08} & \\ 
\\[\dimexpr-\normalbaselineskip+3pt]
27 & $ e + \text{O}_3^{ - }\rightarrow \text{O}_3 + 2e                     $ & $ 3.43 \times 10^{-14} T_e^{0.22} e^{-8.81/T_e} $  &  \cite{Sei03} & \\ 
\\[\dimexpr-\normalbaselineskip+3pt]
28 & $ e + \text{O}_3^{ - } \rightarrow \text{O}_2 + \text{O}(^3P) + 2e           $ & $ 2.69 \times 10^{-14} T_e^{0.06} e^{-20.10/T_e}$  &  \cite{Sei03} & \\ 
\\[\dimexpr-\normalbaselineskip+3pt]
29 & $ e + \text{O}_3^{ - }\rightarrow \text{O}_2 + \text{O}^{ - } + e            $ & $ 3.67 \times 10^{-14} e^{-8.26/T_e} $  &  \cite{Sei03} & \\ 
\\[\dimexpr-\normalbaselineskip+3pt]
30 & $ e + \text{O}_2^+ \rightarrow \text{O}(^3P) + \text{O}(^1D)                 $ & $ 2.20 \times 10^{-14}T_e^{-0.50}     $  &  \cite{Gud04} &   \\ 
\\[\dimexpr-\normalbaselineskip+3pt]
\hline 
\end{tabular}
\label{tab:reo}
\end{table}
\clearpage
	   
\begin{table}\scriptsize
\centering
\caption{Oxygen-Oxygen reactions \cite{Kem15}. The rate coefficients and the 
gas temperature are given in $\text{m}^3\text{s}^{-1}$ and K, respectively.}
\begin{tabular}{p{0.7cm}p{6.5cm}p{5.0cm}p{1.0cm}p{0.2cm}}
\\[\dimexpr-\normalbaselineskip+3pt]
\hline
\\[\dimexpr-\normalbaselineskip+3pt]
\# &Reaction     & Rate Coefficient & Ref & \\ \hline
\\[\dimexpr-\normalbaselineskip+3pt]
31  & $ \text{O}_2^+ + \text{O}^{ - } \rightarrow \text{O}_2 + \text{O}(^3P)                        $ & $ 2.60 \times 10^{-14} (300/T_h )^{0.44} $ &  \cite{Gud04} &   \\ 
\\[\dimexpr-\normalbaselineskip+3pt]
32  & $ \text{O}^+ + \text{O}^{ - } \rightarrow 2\text{O}(^3P)                               $ & $ 4.00 \times 10^{-14} (300/T_h )^{0.43} $ &  \cite{Gud04} &    \\ 
\\[\dimexpr-\normalbaselineskip+3pt]
33  & $ \text{O}(^3P) + \text{O}^{ - } \rightarrow \text{O}_2 + e                            $ & $ 2.30 \times 10^{-16}                   $ &  \cite{Bel05} &    \\ 
\\[\dimexpr-\normalbaselineskip+3pt]
34  & $ \text{O}_2 + \text{O}^+ \rightarrow \text{O}(^3P) + \text{O}_2^+                            $ & $ 2.10 \times 10^{-17} (300/T_h )^{0.50} $ &  \cite{Eli86} &    \\ 
\\[\dimexpr-\normalbaselineskip+3pt]
35  & $ \text{O}_2 + \text{O}(^1D) \rightarrow \text{O}_2 + \text{O}(^3P)                           $ & $ 2.56 \times 10^{-17} e^{67/T_h}      $ &  \cite{Eli86} &    \\ 
\\[\dimexpr-\normalbaselineskip+3pt]
36  & $ \text{O}(^3P) + \text{O}(^1D) \rightarrow 2\text{O}(^3P)                             $ & $ 8.00 \times 10^{-18}                   $ &  \cite{Eli86} &    \\ 
\\[\dimexpr-\normalbaselineskip+3pt]
37  & $ \text{O}_2 (a^1 \Delta_g) + \text{O}^{ - } \rightarrow \text{O}(^3P) + \text{O}_2^{ - }     $ & $ 4.75 \times 10^{-17}                   $ &  \cite{Bel05} & \\ 
\\[\dimexpr-\normalbaselineskip+3pt]
38  & $ \text{O}_2^+ + \text{O}_2^{ - } \rightarrow 2\text{O}_2                              $ & $ 2.01 \times 10^{-13} (300/T_h )^{0.50} $ &  \cite{Eli86} & \\ 
\\[\dimexpr-\normalbaselineskip+3pt]
39  & $ \text{O}^+ + \text{O}_2^{ - } \rightarrow \text{O}_2 + \text{O}(^3P)                        $ & $ 2.70 \times 10^{-13} (300/T_h )^{0.50} $ &  \cite{Kos92} & \\ 
\\[\dimexpr-\normalbaselineskip+3pt]
40 & $ \text{O}(^3P) + \text{O}_2^{ - } \rightarrow \text{O}_2 + \text{O}^{ - }                    $ & $ 3.31 \times 10^{-16}                   $ &  \cite{Eli86} & \\ 
\\[\dimexpr-\normalbaselineskip+3pt]
41 & $ \text{O}_2 (a^1 \Delta_g) + \text{O}_2^{ - } \rightarrow 2\text{O}_2 + e             $ & $ 2.00 \times 10^{-16}                   $ &  \cite{Feh69} & \\ 
\\[\dimexpr-\normalbaselineskip+3pt]
42 & $ \text{O}_2 + \text{O}^{ - } \rightarrow \text{O}_3 + e                               $ & $ 5.00 \times 10^{-21}                   $ &  \cite{Kos92} & \\ 
\\[\dimexpr-\normalbaselineskip+3pt]
43 & $ \text{O}_2 (a^1 \Delta_g) + \text{O}^{ - } \rightarrow \text{O}_3 + e                $ & $ 1.42 \times 10^{-16}                   $ &  \cite{Bel05} & \\ 
\\[\dimexpr-\normalbaselineskip+3pt]
44 & $ \text{O}_3 + \text{O}^+ \rightarrow \text{O}_2 + \text{O}_2^+                               $ & $ 1.00 \times 10^{-16}                   $ &  \cite{Kos92} & \\ 
\\[\dimexpr-\normalbaselineskip+3pt]
45 & $ \text{O}(^3P) + \text{O}_3 \rightarrow 2\text{O}_2                                   $ & $ 1.81 \times 10^{-17} e^{-2300/T_h}   $ &  \cite{Eli86} & \\ 
\\[\dimexpr-\normalbaselineskip+3pt]
46 & $ \text{O}_3 + \text{O}^{ - } \rightarrow \text{O}(^3P) + \text{O}_3^{ - }                    $ & $ 5.30 \times 10^{-16}                   $ &  \cite{Eli86} & \\ 
\\[\dimexpr-\normalbaselineskip+3pt]
47 & $ \text{O}(^3P) + \text{O}_3^{ - } \rightarrow \text{O}_2 + \text{O}_2^{ - }                  $ & $ 1.00 \times 10^{-16}                   $ &  \cite{Eli86} & \\ 
\\[\dimexpr-\normalbaselineskip+3pt]
48 & $ \text{O}(^3P) + \text{O}_3^{ - } \rightarrow 2\text{O}_2 + e                         $ & $ 3.00 \times 10^{-16}                   $ &  \cite{Eli86} & \\ 
\\[\dimexpr-\normalbaselineskip+3pt]
49 & $ \text{O}_2^+ + \text{O}_3^{ - } \rightarrow \text{O}_2 + \text{O}_3                         $ & $ 2.00 \times 10^{-13} (300/T_h )^{0.50} $ &  \cite{Eli86} & \\ 
\\[\dimexpr-\normalbaselineskip+3pt]
50 & $ \text{O}_2^+ + \text{O}_3^{ - } \rightarrow 2\text{O}(^3P) + \text{O}_3                     $ & $ 1.01 \times 10^{-13} (300/T_h )^{0.50} $ &  \cite{Eli86} & \\ 
\\[\dimexpr-\normalbaselineskip+3pt]
51 & $ \text{O}_3 + \text{O}_2^{ - } \rightarrow \text{O}_2 + \text{O}_3^{ - }                     $ & $ 4.00 \times 10^{-16}                   $ &  \cite{Eli86} & \\ 
\\[\dimexpr-\normalbaselineskip+3pt]
52 & $ \text{O}(^3P) + \text{O}_2^{ - } \rightarrow \text{O}_3 + e                          $ & $ 3.30 \times 10^{-16}                   $ &  \cite{Feh67} & \\ 
\\[\dimexpr-\normalbaselineskip+3pt]
53 & $ \text{O}_2 + \text{O}(^1D) \rightarrow \text{O}_2 (a^1 \Delta_g) + \text{O}(^3P)            $ & $ 1.00 \times 10^{-18}                   $ &  \cite{Eli86} &    \\ 
\\[\dimexpr-\normalbaselineskip+3pt]
54 & $ \text{O}_2 (b^1 \sum_g^+) + \text{O}^{ - } \rightarrow \text{O}_2 + \text{O}(^3P) + e       $ & $ 6.90 \times 10^{-16}                   $ &  \cite{Ale78} &    \\ 
\\[\dimexpr-\normalbaselineskip+3pt]
55 & $ \text{O}_2 (b^1 \sum_g^+) + \text{O}(^3P) \rightarrow \text{O}_2 (a^1 \Delta_g) + \text{O}(^3P) $ & $ 8.10 \times 10^{-20}                   $ &  \cite{Gor95} &    \\ 
\\[\dimexpr-\normalbaselineskip+3pt]
56 & $ \text{O}_2 + \text{O}_2 (b^1 \sum_g^+) \rightarrow \text{O}_2 + \text{O}_2 (a^1 \Delta_g)       $ & $ 3.79 \times 10^{-22} (300/T_h )^{-2.40} e^{-281/T_h} $ &  \cite{Gor95}  &    \\ 
\\[\dimexpr-\normalbaselineskip+3pt]
57 & $ \text{O}_2 (a^1 \Delta_g) + \text{O}(^3P) \rightarrow \text{O}_2 + \text{O}(^3P)                $ & $ 1.30 \times 10^{-22}                   $ &  \cite{Cla69} &    \\ 
\\[\dimexpr-\normalbaselineskip+3pt]
58 & $ \text{O}_2^+ + \text{O}^{ - } \rightarrow 3\text{O}(^3P)                                 $ & $ 2.60 \times 10^{-14} (300/T_h )^{0.44} $ &  \cite{Gud04} &    \\ 
\\[\dimexpr-\normalbaselineskip+3pt]
59 & $ \text{O}_2 + \text{O}_2 (a^1 \Delta_g) \rightarrow 2\text{O}_2                           $ & $ 2.20 \times 10^{-24} (300/T_h )^{-0.80}$ &  \cite{Kos92} &    \\ 
\\[\dimexpr-\normalbaselineskip+3pt]
60 & $ \text{O}_2 (A^3 \sum_u^+,...) + \text{O}(^3P) \rightarrow \text{O}_2 (b^1 \sum_g^+) + \text{O}(^1D) $ & $ 1.35 \times 10^{-18} $ &  \cite{Vas04} &    \\ 
\\[\dimexpr-\normalbaselineskip+3pt]
61 & $ \text{O}_2 + \text{O}_3 \rightarrow 2\text{O}_2 + \text{O}(^3P)                                          $ & $ 7.26 \times 10^{-16} e^{-11400/T_h}  $ &  \cite{Eli86} & \\ 
\\[\dimexpr-\normalbaselineskip+3pt]
62 & $ \text{O}_2 (a^1 \Delta_g) + \text{O}_2 (a^1 \Delta_g) \rightarrow \text{O}_2 + \text{O}_2 (b^1 \sum_g^+)$ & $ 1.80 \times 10^{-24} (300/T_h )^{-3.80} e^{700/T_h} $ &  \cite{Coh83} &    \\ 
\\[\dimexpr-\normalbaselineskip+3pt]
63 & $ \text{O}_2 (a^1 \Delta_g) + \text{O}_2 (a^1 \Delta_g) \rightarrow 2\text{O}_2                    $ & $ 5.50 \times 10^{-29} (300/T_h )^{-0.50} $ &  \cite{Bor83} &    \\ 
\\[\dimexpr-\normalbaselineskip+3pt]
64 & $ \text{O}_2 (b^1 \sum_g^+) + \text{O}_3 \rightarrow 2\text{O}_2 + \text{O}(^3P)                          $ & $ 1.50 \times 10^{-17}                    $ &  \cite{Jeo00} & \\ 
\\[\dimexpr-\normalbaselineskip+3pt]
65 & $ \text{O}_2 (a^1 \Delta_g) + \text{O}_3 \rightarrow 2\text{O}_2 + \text{O}(^3P)                          $ & $ 6.01 \times 10^{-17} e^{-2853/T_h}    $ &  \cite{Jeo00} & \\ 
\\[\dimexpr-\normalbaselineskip+3pt]
66 & $ \text{O}_2 + \text{O}_2 (A^3 \sum_u^+,...)\rightarrow2 \text{O}_2 (b^1 \sum_g^+) $ & $ 2.90 \times 10^{-19}  $ &  \cite{Ken80} &    \\ 
\\[\dimexpr-\normalbaselineskip+3pt]
67 & $ \text{O}_2^+ + \text{O}_2^{ - } \rightarrow \text{O}_2 + 2\text{O}(^3P)                                        $ & $ 1.01 \times 10^{-13} (300/T_h )^{0.50}  $ &  \cite{Eli86} & \\ 
\\[\dimexpr-\normalbaselineskip+3pt]
68 & $ \text{O}_2 (A^3 \sum_u^+,...) \rightarrow \text{O}_2 + \hbar \nu       $ & $ 6.25 1/s                        $ &  \cite{Ken83} &    \\ 
\\[\dimexpr-\normalbaselineskip+3pt]
69 & $ \text{O}_2 (A^3 \sum_u^+,...) + \text{O}(^3P) \rightarrow \text{O}_2 + \text{O}(^3P) $ & $ 4.95 \times 10^{-18}                    $ &  \cite{Vas04} &    \\ 
\\[\dimexpr-\normalbaselineskip+3pt]
70 & $ \text{O}_2 (A^3 \sum_u^+,...) + \text{O}(^3P) \rightarrow \text{O}_2 (a^1 \Delta_g) + \text{O}(^1D) $ & $ 2.70 \times 10^{-18}   $ &  \cite{Vas04} &    \\ 
\\[\dimexpr-\normalbaselineskip+3pt]
71 & $ 2\text{O}_2 (b^1 \sum_g^+) \rightarrow \text{O}_2 + \text{O}_2 (a^1 \Delta_g)       $ & $ 3.60 \times 10^{-23} (300/T_h )^{-0.50}  $ &  \cite{Sta04} &    \\ 
\\[\dimexpr-\normalbaselineskip+3pt]
\hline 
\end{tabular}
\label{tab:roo}
\end{table}
\clearpage

\begin{table}\scriptsize
\centering
\caption{Three body reactions \cite{Kem15}. The rate coefficients and the gas 
temperature are given in $\text{m}^6\text{s}^{-1}$ and K, respectively, whereas
the electron temperature is in eV. }
\begin{tabular}{p{0.7cm}p{6.5cm}p{5.0cm}p{1.0cm}p{0.2cm}}
\\[\dimexpr-\normalbaselineskip+3pt]
\hline
\\[\dimexpr-\normalbaselineskip+3pt]
\# &Reaction     & Rate Coefficient & Ref & \\ \hline
\\[\dimexpr-\normalbaselineskip+3pt]
72  & $ e + e + \text{O}^+ \rightarrow \text{O}(^3P) + e              $ & $ 7.89 \times 10^{-39} T_e^{-4.50}  $ &  \cite{Eli86} &    \\ 
\\[\dimexpr-\normalbaselineskip+3pt]
73  & $ e + \text{O}_2 + \text{O}_2 \rightarrow \text{O}_2 + \text{O}_2^{ - }         $ & $ 2.26 \times 10^{-42} T_e^{0.50}   $ &  \cite{Shi81} & \\ 
\\[\dimexpr-\normalbaselineskip+3pt]
74  & $ \text{O} + \text{O}_2 + \text{O}_2  \rightarrow \text{O}_3 + \text{O}_2        $ & $ 6.3 \times 10^{-46} (300/T_h)^2   $ &  \cite{Lop11} & \\ 
\\[\dimexpr-\normalbaselineskip+3pt]
75  & $ \text{O}_2 + \text{O}(^3P) + \text{O}(^3P) \rightarrow \text{O}(^3P) + \text{O}_3  $ & $ 2.15 \times 10^{-40} e^{345/T_h}   $ &  \cite{Tho2010} & \\ 
\\[\dimexpr-\normalbaselineskip+3pt]
76  & $ e + \text{O}_2 + \text{O}(^3P) \rightarrow \text{O}(^3P) + \text{O}_2^{ - }   $ & $ 1.00 \times 10^{-43}                 $ &  \cite{Eli86} & \\ 
\\[\dimexpr-\normalbaselineskip+3pt]
77  & $ e + \text{O}_2 + \text{O}(^3P) \rightarrow \text{O}_2 + \text{O}^{ - }      $ & $ 1.00 \times 10^{-43}                 $ &  \cite{Eli86} &    \\ 
\\[\dimexpr-\normalbaselineskip+3pt]
78  & $ e + \text{O}_2 + \text{O}^+ \rightarrow \text{O}_2 + \text{O}(^3P)          $ & $ 1.00 \times 10^{-38}                 $ &  \cite{Eli86} &    \\ 
\\[\dimexpr-\normalbaselineskip+3pt]
79  & $ \text{O}_2 + \text{O}_2 + \text{O}^{ - } \rightarrow \text{O}_2 + \text{O}_3^{ - }  $ & $ 1.00 \times 10^{-42} (300/T_h)       $ &  \cite{Eli86} & \\ 
\\[\dimexpr-\normalbaselineskip+3pt]
80  & $ \text{O}_2 + \text{O}^+ + \text{O}^{ - } \rightarrow \text{O}_2 + \text{O}_2       $ & $ 2.10 \times 10^{-37} (300/T_h)^{2.50}$ &  \cite{Eli86} &    \\ 
\\[\dimexpr-\normalbaselineskip+3pt]
81 & $ \text{O}_2 + \text{O}_2^+ + \text{O}^{ - } \rightarrow \text{O}_2 + \text{O}_3       $ & $ 2.01 \times 10^{-37} (300/T_h)^{2.50}$ &  \cite{Eli86} & \\ 
\\[\dimexpr-\normalbaselineskip+3pt]
82 & $ \text{O}_2 + \text{O}_2 + \text{O}(^3P) \rightarrow \text{O}_2 + \text{O}_3        $ & $ 6.90 \times 10^{-46} (300/T_h)^{1.25}$ &  \cite{Eli86} & \\ 
\\[\dimexpr-\normalbaselineskip+3pt]
83 & $ \text{O}_2 + \text{O}(^3P) + \text{O}(^3P) \rightarrow \text{O}_2 + \text{O}_2 (A^3 \sum_u^+,...) $ & $ 1.20 \times 10^{-46} $ &  \cite{Ken80} &    \\ 
\\[\dimexpr-\normalbaselineskip+3pt]
84 & $ \text{O}_2 + \text{O}_2 (a^1 \Delta_g) + \text{O}(^3P) \rightarrow \text{O}_2 + \text{O}_2 + \text{O}(^3P)$ & $ 1.00 \times 10^{-44} $ &  \cite{Vas04} &    \\ 
\\[\dimexpr-\normalbaselineskip+3pt]
85 & $ 3\text{O}(^3P) \rightarrow \text{O}_2 (a^1 \Delta_g) + \text{O}(^3P)                 $ & $ 1.93 \times 10^{-47} (300/T_h)^{0.63}$ &  \cite{Gor95} &    \\ 
\\[\dimexpr-\normalbaselineskip+3pt]
86 & $ \text{O}_2 + \text{O}(^3P) + \text{O}(^3 P ) \rightarrow \text{O}_2 + \text{O}_2 (a^1 \Delta_g)    $ & $ 6.93 \times 10^{-47} (300/T_h)^{0.63}$ &  \cite{Gor95} &    \\ 
\\[\dimexpr-\normalbaselineskip+3pt]
\hline 
\end{tabular}
\label{tab:rtb}
\end{table}

\begin{table}\scriptsize
\centering
\caption{Elastic electronic collisions included in the simulations.}
\begin{tabular}{p{0.3cm}p{2cm}p{0.5cm}p{0.5cm}}
\\[\dimexpr-\normalbaselineskip+3pt]
\hline
\\[\dimexpr-\normalbaselineskip+3pt]
\# &Collision     &  Ref  & \\ \hline
\\[\dimexpr-\normalbaselineskip+3pt]
1  & $ e+ \text{Ar}    $ &  \cite{Gud02_report} &      \\ 
\\[\dimexpr-\normalbaselineskip+3pt]
2  & $ e+ \text{O}_2     $ &  \cite{Iti09} &      \\ 
\\[\dimexpr-\normalbaselineskip+3pt]
3  & $  e+ \text{O}    $ &  \cite{Iti90}   &     \\ 
\\[\dimexpr-\normalbaselineskip+3pt]
4  & $  e+\text{O}_3    $ &  \cite{Gup05}  &     \\ 
\\[\dimexpr-\normalbaselineskip+3pt]
\hline 
\end{tabular}
\label{tab:ela}
\end{table}

\begin{table}\scriptsize
\centering
\caption{Chemical reactions induced at the wall.}
\label{tab:wall}
\begin{tabular}{llcccl}
\\[\dimexpr-\normalbaselineskip+3pt]
\hline
\\[\dimexpr-\normalbaselineskip+3pt]
\# &Reaction     &  probability($\gamma$) $R_p$ & probability($\gamma$) $R$ & Ref &  \\ \hline
\\[\dimexpr-\normalbaselineskip+3pt]
1  & $ \text{Ar}(4s,4p) + \text{wall} \rightarrow \text{Ar}  $ & $1$ & $1$ &  \cite{Jim12} &    \\ 
\\[\dimexpr-\normalbaselineskip+3pt]
1  & $ \text{O}(^3P,^1D) + \text{wall} \rightarrow 1/2\text{O}_2  $ & $0.09$ & $0.09$ &  \cite{Sta08} &    \\ 
\\[\dimexpr-\normalbaselineskip+3pt]
2  & $ \text{O}(^1D) + \text{wall} \rightarrow \text{O}(^3P)  $ & $0.1$  & $0.1$  &  \cite{Tho2010} &    \\ 
\\[\dimexpr-\normalbaselineskip+3pt]
3  & $ \text{O}_2(a^1 \Delta_g) + \text{wall} \rightarrow \text{O}_2  $ & $0.007$ & $0.007$ & \cite{Sha89} &     \\ 
\\[\dimexpr-\normalbaselineskip+3pt]
4  & $ \text{O}_2(b^1 \sum_g^+) + \text{wall} \rightarrow \text{O}_2  $ & $0.1$  & $0.1$  & \cite{Tho2010} &     \\ 
\\[\dimexpr-\normalbaselineskip+3pt]
5  & $ \text{O}_2(A^3 \sum_u^+,...) + \text{wall} \rightarrow \text{O}_2  $ & $0.1$  & $0.1$  &  \cite{Tho2010} &    \\ 
\\[\dimexpr-\normalbaselineskip+3pt]
\hline
\end{tabular}
\label{tab:wr}
\end{table}
\clearpage
\twocolumngrid